\documentclass[times,final]{elsarticle}

\graphicspath{{figs/}}
\usepackage{siunitx}  
\usepackage{subcaption}
\usepackage{booktabs}
\usepackage{placeins}

\usepackage{acro}

\DeclareAcronym{2afc}{
short = 2AFC,
long = two-alternative forced choice,
}

\DeclareAcronym{hvs}{
short = HVS,
long = human visual system,
}

\DeclareAcronym{sph}{
short = SPH,
short-indefinite = an,
long = smoothed particle hydrodynamics,
}

\DeclareAcronym{flip}{
short = FLIP,
long = fluid-implicit-particle,
}

\DeclareAcronym{simple}{
short = SIMPLE,
long = semi-implicit method for pressure linked equations,
}

\DeclareAcronym{eno}{
short = ENO,
long = essentially non-oscillatory,
}

\DeclareAcronym{rmse}{
short = RMSE,
long = root-mean-square error,
}

\DeclareAcronym{psnr}{
short = PSNR,
long = peak signal-to-noise ratio,
}

\DeclareAcronym{ssim}{
short = SSIM,
long = structure similarity metric,
}

\DeclareAcronym{tgv}{
short = TGV,
long = Taylor-Green vortex,
}

\DeclareAcronym{ncm}{
short = NCM,
long = near-convergence consistency metric,
}

\newcommand{\maxhi}[1]{\underline{\textbf{#1}}}
\newcommand{\myeqref}[1]{Eq.~(\ref{#1})}
\newcommand{\myfigref}[1]{Fig.~\ref{#1}}
\newcommand{\myfigsref}[2]{Fig.~\ref{#1}--\ref{#2}}
\newcommand{\mytabref}[1]{Table~\ref{#1}}
\newcommand{\mycite}[1]{\citep{#1}}

\usepackage[normalem]{ulem}     
\usepackage{color}

\definecolor{R}{rgb}{1, 0, 0}
\definecolor{G}{rgb}{0, 0.5, 0}
\definecolor{B}{rgb}{0, 0, 1}

\definecolor{kwumCol}{rgb}{0.9, 0, 0.2}
\definecolor{nilsCol}{rgb}{1., 0.4, 0.}
\definecolor{xyhuCol}{rgb}{0.9, 0, 0.9}

\usepackage{jcomp4arxiv}
\usepackage{framed,multirow}

\usepackage{amssymb}
\usepackage{latexsym}

\usepackage{url}
\usepackage{xcolor}
\definecolor{newcolor}{rgb}{.8,.349,.1}

\journal{Journal of Computational Physics}

\begin{document}

\verso{Um \textit{et al.}}

\begin{frontmatter}

\title{Spot the Difference: Accuracy of Numerical Simulations via the Human Visual System}


\author[1]{Kiwon Um}
\ead{kiwon.um@tum.de}
\author[1]{Xiangyu Hu}
\ead{xiangyu.hu@tum.de}
\author[2]{Bing Wang}
\ead{wbing@tsinghua.edu.cn}
\author[1]{Nils Thuerey\corref{cor1}}
\ead{nils.thuerey@tum.de}
\cortext[cor1]{Corresponding author.}

\address[1]{Technical University of Munich, Garching, 85748, Germany}
\address[2]{Tsinghua University, Beijing, 100084, China}




\begin{abstract}
  Comparative evaluation lies at the heart of science, and determining the
  accuracy of a computational method is crucial for evaluating its potential as
  well as for guiding future efforts. However, metrics that are typically used
  have inherent shortcomings when faced with the under-resolved solutions of
  real-world simulation problems. We show how to leverage crowd-sourced user
  studies in order to address the fundamental problems of widely used classical
  evaluation metrics. We demonstrate that such user studies, which inherently
  rely on the human visual system, yield a very robust metric and consistent
  answers for complex phenomena without any requirements for proficiency
  regarding the physics at hand. This holds even for cases away from convergence
  where traditional metrics often end up inconclusive results. More
  specifically, we evaluate results of different \ac{eno} schemes in different
  fluid flow settings. Our methodology represents a novel and practical approach
  for scientific evaluations that can give answers for previously unsolved
  problems.
\end{abstract}


\end{frontmatter}

\section{Introduction}

Evaluation is an essential process for any form of scientific research, and
numerical simulations are no exception. On the contrary, being able to reliably
evaluate the quality of an output from a numerical method is crucial to judge
its usefulness and to steer future efforts for developing new methods
\mycite{montecinos2012,johnsen2011,peshkov2019,zhao2019}. Traditionally, the
evaluation of numerical results relies on simple metrics in the form of vector
norms \mycite{christie2005,press2007,kat2012} to compute the distance between
the approximation and a ground truth target. The latter can take the form of an
analytic solution or can be obtained with a highly refined numerical solution
that yields a converged result \mycite{oberkampf2004}. A large body of work on
this topic exists, and it has been particularly important for all forms of
transient fluid flow problems, which encompass a vast number of important
research questions, including classic airfoil flows \mycite{rhie1983numerical},
cavitation problems \mycite{stieger2017}, and the challenges of turbulence
\mycite{serra2019,smyth2019}.

Despite the widespread use and huge impact of flow simulations on many
real-world problems, evaluating the quality of a result remains an extremely
challenging and open problem. The non-linear nature of the underlying equations
leads to solutions for which slight changes in boundary conditions or simulation
parameters lead to very significant and unpredictable changes in the output
\mycite{johnsen2010}. For such non-trivial cases, vector norms (e.g., the $l^2$
norm), which are commonly used to quantify differences of solutions, were shown
to be highly unreliable \mycite{wang2006}. Although these problems are widely
known \mycite{mehta2016}, no established alternatives exist. As a consequence,
the resulting evaluations can fail to effectively discriminate different sets of
results or methods.

In order to address these central challenges of performance evaluations, we
propose to employ the \ac{hvs}, which is widely regarded to be particularly
powerful to analyze and compare visual data. People even enjoy challenging their
visual system in games such as \emph{Spot the Difference}, and a large body of
scientific work has established the fact that humans excel at assessing the
similarity and differences of images \mycite{cornsweet1970,neri2002}. The
\ac{hvs} has also demonstrated its ability to support science projects via
crowdsourcing, e.g., for experimental behavioral research \mycite{crump2013},
scoring chemistry images \mycite{irshad2017}, and astronomical discovery
\mycite{spacewarps}. We will demonstrate that carefully designed crowdsourcing
studies with a reference solution make it possible to reliably rank simulation
results and, in this way, obtain a relative quantification of the accuracy of
each result. Interestingly, no expert knowledge and cognition are required from
user study participants \mycite{albright2002}. We demonstrate with a large-scale
series of studies that non-professionals can provide reliable and robust
evaluations in fields, which previously were the sole domain of experts
\mycite{larkin1980}.

\section{Evaluation methodology based on user studies}

\begin{figure}[tbh]
  \centering
  \def\svgwidth{\linewidth}
  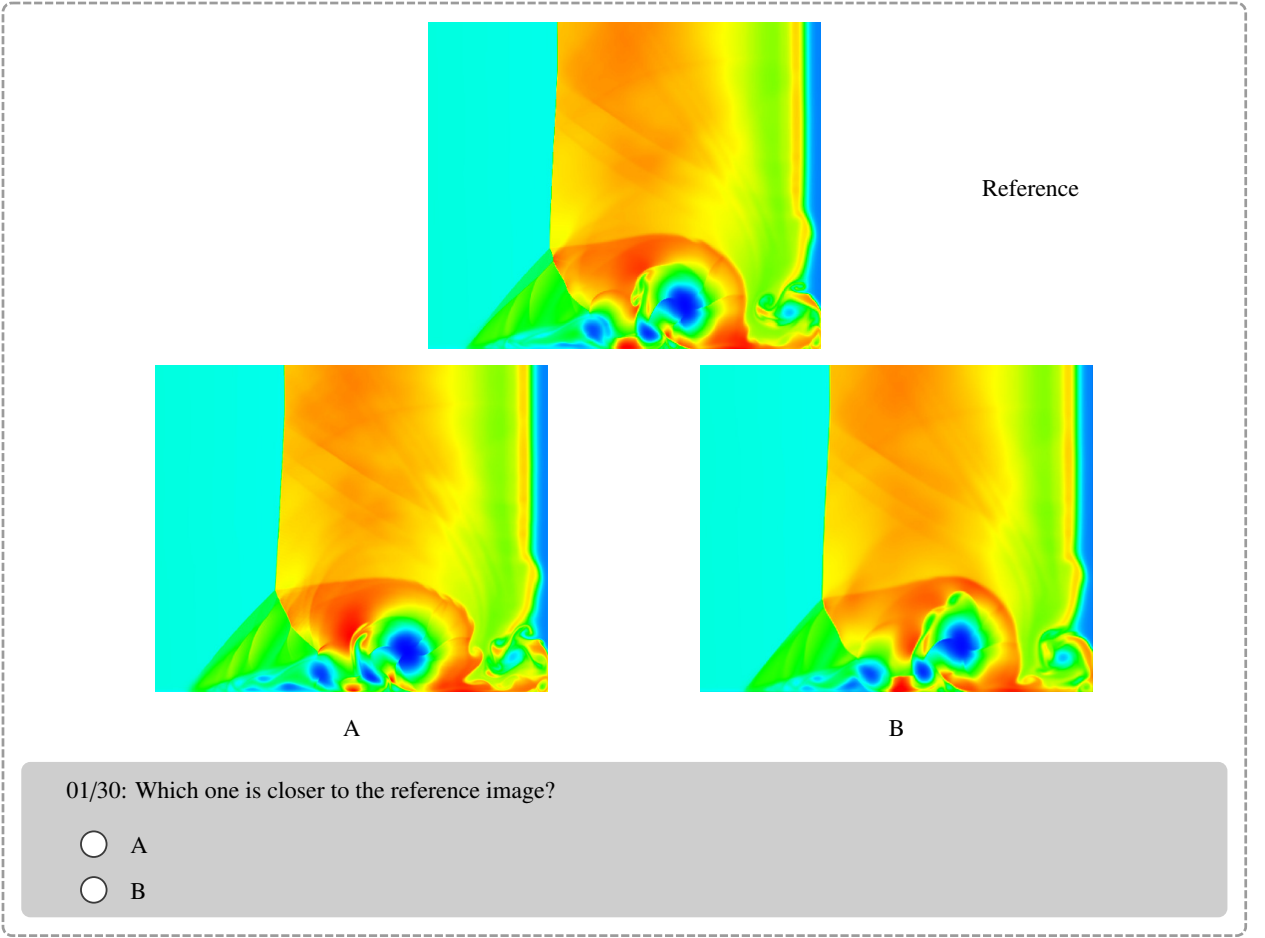
  \caption{User study design: All user studies in this paper were conducted with
    the design shown above. Each participant has to make one choice between A
    and B per question.}
  \label{fig:study}
\end{figure}

The design goal for our user studies is to make the task as simple as possible
in order to minimize misunderstandings on the side of the participants. This
reduces noise and makes the answers more reliable, i.e., meaningful. This goal
motivates the use of the \ac{2afc} design \mycite{fechner1860} where the
participants are given two options and have to choose one of them. In our case,
both options are given as images, and the participants are shown a third image
as reference with the task to identify which one of the options is closer to the
reference. Intentionally, no explanation of the image content or further
instruction regarding which features in the images to focus on is given. Hence,
in the limit of large enough numbers of randomly selected participants, the
evaluations are expected to follow the natural human perception. See
\myfigref{fig:study} for an example task of our study.

\subsection{Performance score}\label{sec:score}

After collecting the votes for a set of pairwise comparisons among $N$
candidates (i.e., images) from a user study, we transform the participants'
pairwise answers into a set of scores, $\mathbf{s}=\{s_1, s_2, ..., s_N\}$,
using the Bradley-Terry model \mycite{bradley1952}. Let $p_{ij}$ be the
probability that a participant chooses image $i$ over image $j$:
\begin{equation}
  \label{eq:p-i-beat-j}
  p_{ij} = \frac{e^{s_i - s_j}}{1 + e^{s_i - s_j}}.
\end{equation}
Denoting the number of votes where the participants chose image $i$ over image
$j$ as $w_{ij}$ and assuming each vote is independent, we can represent the log
likelihood for all pairs among all candidates:
\begin{equation}
  \label{eq:bt-likelihood}
  L(\mathbf{s}) = \sum_{i=1}^m\sum_{j=1}^m \left(w_{ij}s_i - w_{ij} \ln(e^{s_i} + e^{s_j})\right).
\end{equation}
We then compute the scores of all images by maximizing the likelihood $L$ of
\myeqref{eq:bt-likelihood} with respect to all votes collected from the user
study \mycite{hunter2004}. The set of scores $\mathbf{s}$ then represents how
perceptually similar to the reference each candidate is, and differences in
scores between two candidates can be used to compute $p_{ij}$ with
\myeqref{eq:p-i-beat-j}. Due to the statistical nature of user study
measurements, each evaluation is associated with a confidence interval, which in
practice can be narrowed with increased participant numbers if necessary.

\subsection{Winning probability and \acl{ncm}}\label{sec:index}

Once we have computed the scores $s_i$ for all $N$ candidates considered in a
study, we can also calculate the probability that a candidate $i$ is chosen
among all candidates. This \emph{winning probability} is given by the average
probability of $p_{ij}$ that a participant chooses the candidate $i$ over the
others:
\begin{equation}
  \label{eq:win-index}
  W_i = \frac{1}{N-1}\sum_{j \neq i}^N p_{ij} = \frac{1}{N-1}\sum_{j \neq i}^N \frac{e^{s_i - s_j}}{1+e^{s_i - s_j}}.
\end{equation}
In contrast to the pairwise probability $p_{ij}$, $W_i$ represents the
probability that a candidate $i$ was preferred over all other candidates.

Calculating the winning probabilities for one candidate across different
studies, we can define its \emph{\acl{ncm}} $\varepsilon_i$ as the standard
deviation of its winning probabilities:
\begin{equation}
  \label{eq:ncm}
  \varepsilon_i = \sqrt{\frac{1}{M-1}\sum_{j=1}^{M}\left(W_{i,j} - \mu_i \right)^2}
\end{equation}
where $M$ denotes the number of different studies and $\mu$ denotes the mean
winning probability. This metric indicates how consistently a candidate performs
over the others across different evaluation studies.

\section{Simulation setups}

Our evaluation primarily focuses on seven numerical methods that belong to the
class of \ac{eno} schemes \mycite{harten1986}. They represent a widely used
class of finite difference schemes in computational fluid dynamics
\mycite{dumbser2007arbitrary,balsara2009efficient,johnsen2011} and rely on
multiple interpolation stencils where each defines a candidate approximation.
These candidates are selected individually or combined together to obtain the
final result. Since the criteria for selecting or combining them are based on
smoothness of these stencils, the final interpolation is able to achieve good
numerical stability if the criteria are biased towards smoothness. One of the
most popular variations of this scheme is the \emph{weighted} \ac{eno} scheme
\mycite{jiang1996}, which is denoted by W5 above. Here, the final interpolation
$q^{l}_{i+1/2}$ is computed with the weighted average:
\begin{equation}
  \label{final-weighting}
  q^{l}_{i+1/2} = \sum_{k=1}^m{w_k}\tilde{q}_{i+1/2}^{k}
\end{equation}
where $w_k$ and $\tilde q_{i+1/2}^{k}$ are the non-linear weight and the
approximation obtained from the candidate stencil $k$, respectively. The weights
are computed as follows:
\begin{equation}
  \label{weight-new}
  \omega_{k} = \frac{\alpha_{k}}{\sum^{m}_{s=1}\alpha_{s}} \quad \mathrm{and}
  \quad \alpha_{k} = \frac{d_k}{\beta^2_{k} + \epsilon}
\end{equation}
where $d_k$ is the optimal weight, $\beta_k$ is the smoothness indicator for the
candidate stencil, and $\epsilon$ is a small positive number. The optimal weight
$d_k$ is chosen in such a way that the final interpolation achieves higher order
of accuracy than that of the candidate stencil.

In addition to W5, we chose two further modifications: W5z \mycite{borges2008}
and W6c \mycite{hu2010}, which are representatives of improved weighted \ac{eno}
schemes. Besides, we also consider four \emph{targeted} \ac{eno} schemes
\mycite{fu2016}: T5, T5o, T6, and T6o, which are more recent variations that are
motivated by the stencil-selection strategy of the original \ac{eno} scheme. In
our notation, the number 5 or 6 indicates the size of full discretization
stencil, which includes all points of the candidate stencil, and also represents
the order of accuracy in the smoothing region.

In the following, we describe six simulation setups used in our evaluations. We
include three different setups using the \ac{eno} variants mentioned above,
which will be evaluated in detail in terms of their relative performance. Below,
we also describe an additional set of three fluid flow solvers that will be used
for evaluating the general robustness of our user study approach.

\subsection{Viscous shock tube}
\begin{figure}
  \captionsetup[subfigure]{aboveskip=1pt,belowskip=3pt}
  \centering
  \subcaptionbox*{1$\times$} {\includegraphics[width=0.325\linewidth]{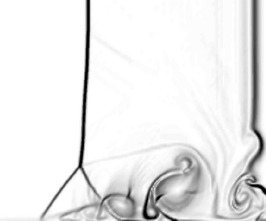}}
  \subcaptionbox*{2$\times$} {\includegraphics[width=0.325\linewidth]{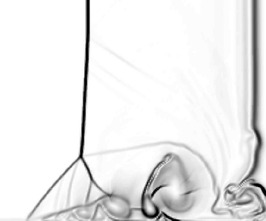}}
  \subcaptionbox*{4$\times$} {\includegraphics[width=0.325\linewidth]{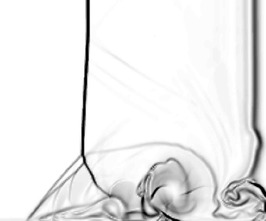}}\\
  \subcaptionbox*{8$\times$} {\includegraphics[width=0.325\linewidth]{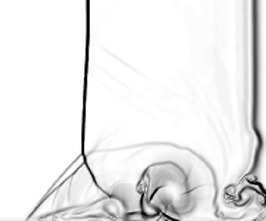}}
  \subcaptionbox*{16$\times$}{\includegraphics[width=0.325\linewidth]{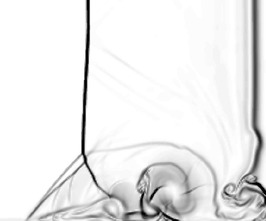}}
  \subcaptionbox*{Reference} {\includegraphics[width=0.325\linewidth]{figs/exp.vshock/ref}}\\
  \caption{Viscous shock tube: This setup contains simulations of a complex
    unsteady viscous shock flow \mycite{daru2000} with different resolutions. The
    reference for the user studies was given as the highest resolution result,
    i.e., the one with the smallest discretization error. Due to the simulated
    viscosity, the sequence converges towards a result very similar to the
    ground truth.}
  \label{fig:viscous-shock-res}
\end{figure}

The flow solver for this case follows the general \ac{eno} methodology
\mycite{harten1986} except that the interpolation is applied with a W5 scheme.
Specifically, the Roe approximation is used for the characteristic decomposition
at the computational cell faces, the Lax-Friedrichs formulation is used for the
numerical fluxes, and the 3rd-order Total Variation Diminishing Runge–Kutta
scheme is used for time integration \mycite{shu1988efficient}. In this
two-dimensional case, the shock wave Mach number is 2.37
\mycite{pirozzoli2006direct}, the viscosity is assumed to be constant, and the
Prandtl number is 0.73, which gives a Reynolds number of 1000 \mycite{fu2016}.
The simulations are performed on uniform grids of 320$\times$160 (1$\times$),
640$\times$320 (2$\times$), 1280$\times$640 (4$\times$), 2560$\times$1280
(8$\times$), and 5120$\times$2560 (16$\times$). The simulations are visualized
with the density gradient sampled on the lowest resolution as shown in
\myfigref{fig:viscous-shock-res}, while the reference visualizes the data set of
16$\times$ without down-sampling.

\subsection{Double Mach reflection}
\begin{figure}
  \captionsetup[subfigure]{aboveskip=1pt,belowskip=3pt}
  \centering
  \subcaptionbox*{1$\times$} {\includegraphics[width=0.325\linewidth]{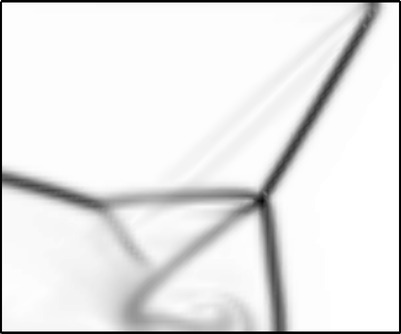}}
  \subcaptionbox*{2$\times$} {\includegraphics[width=0.325\linewidth]{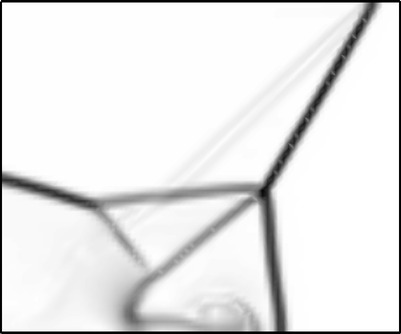}}
  \subcaptionbox*{4$\times$} {\includegraphics[width=0.325\linewidth]{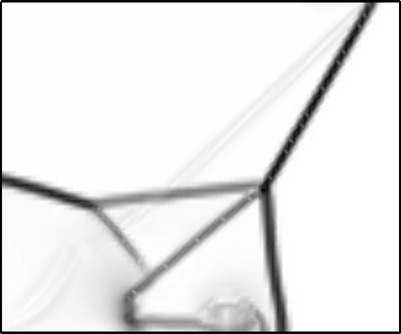}}\\
  \subcaptionbox*{8$\times$} {\includegraphics[width=0.325\linewidth]{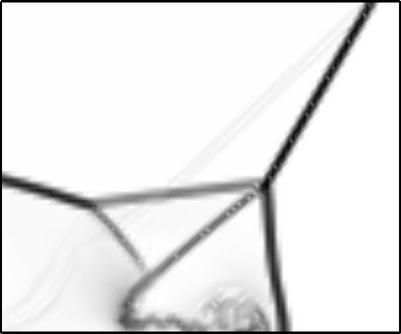}}
  \subcaptionbox*{16$\times$}{\includegraphics[width=0.325\linewidth]{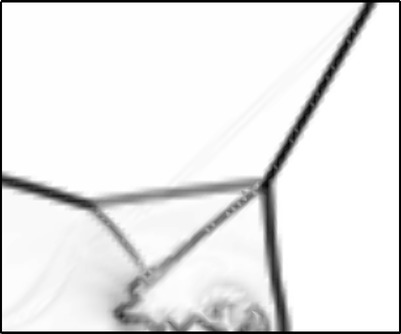}}
  \subcaptionbox*{Reference} {\includegraphics[width=0.325\linewidth]{figs/exp.ishock/ref}}\\
  \caption{Double Mach reflection: This setup contains simulations of a
    two-dimensional inviscid flow with a strong shock \mycite{woodward1984} with
    different resolutions. The reference data set was the highest resolution
    result. Being an inviscid flow problem, this sequence does not exhibit
    numerical convergence in the traditional sense. Despite this, participants
    of the corresponding user study had no significant problems establishing the
    correct ordering.}
  \label{fig:inviscid-shock-res}
\end{figure}

The underlying flow solver is the same as that for the viscous shock tube above;
but, it solves the inviscid Euler equation instead. In this two-dimensional
case, a right-moving Mach 10 shock wave is reflected by a horizontal wall with
an incidence angle of 60$^o$ \mycite{jiang1996}. The simulations are done on
uniform grids of 240$\times$60 (1$\times$), 480$\times$120 (2$\times$),
960$\times$240 (4$\times$), 1920$\times$480 (8$\times$), and 3840$\times$960
(16$\times$). The simulations are visualized with the density gradient sampled
on the lowest resolution, as shown in \myfigref{fig:inviscid-shock-res}, while
the reference visualizes the data set of 16$\times$ without down-sampling.

\subsection{Taylor-Green vortex}
Although the \ac{tgv} \mycite{taylorGreen,brachet1983} is a typical
incompressible flow problem, the solver used here is the same as above used for
simulating compressible flows with shocks, except different \ac{eno} schemes are
applied. Since the Mach number is less than 0.1, the density variations are
negligible in this case. The Reynolds number of the problem is 3600. As a
consequence, turbulent flow develops shortly after the initial condition. The
simulations are carried out in a periodic box on uniform grids of 64$^3$
(1$\times$), 128$^3$ (2$\times$), and 256$^3$ (4$\times$). The reference
simulation is done on a grid of 512$^3$ (8$\times$). The simulations are
visualized according to the Q-criterion \mycite{hunt1988eddies} for an
isosurface of $Q=3$ and colored with the $x$-component of vorticity as shown in
\myfigref{fig:tgv:all}.

Next, we describe the three solver variants that will be used for a more general
evaluation of our user study methodology below.

\subsection{Breaking dam}
The first of the three setups used to evaluate the robustness of our user
studies employs two representative Eulerian and Lagrangian free surface liquid
simulation methods: \ac{flip} \mycite{brackbill1988,zhu2005} and \ac{sph}
\mycite{monaghan2005,adami2012}. The top left of \myfigref{fig:dam} shows the
simulation setup \mycite{kleefsman2005}. \ac{flip} uses a regular grid of
80$\times$75$\times$25 for (1$\times$), 160$\times$150$\times$50 for
(2$\times$), and 320$\times$330$\times$100 for (4$\times$) while sampling each
cell using 2$^3$ particles; i.e., each simulation uses 83k, 664k, and 5,315k
particles, respectively. Likewise, \ac{sph} uses 80k Lagrangian particles for
(1$\times$), 665k for (2$\times$), and 2,253k for (3$\times$). Both methods
simulate this setup with the gravity of 9.8\si{m}/\si{s} and without additional
viscosity and surface tension. \ac{flip} uses first-order boundary conditions
for the Dirichlet pressure boundary conditions at the free surface and free-slip
velocity boundary conditions for solid surfaces. The \ac{sph} solver is based on
the work of \mycite{adami2012}. We use the cubic spline kernel for the \ac{sph}
quantities and dummy particles at the wall for boundary conditions of pressure
and density.

\subsection{Plume}
\begin{figure}
  \captionsetup[subfigure]{aboveskip=1pt,belowskip=3pt}
  \centering
  \subcaptionbox*{Reference}   {\includegraphics[width=0.24\linewidth]{figs/exp.plume/out_n00000_n05_n0100}}
  \subcaptionbox*{1$\times$}   {\includegraphics[width=0.24\linewidth]{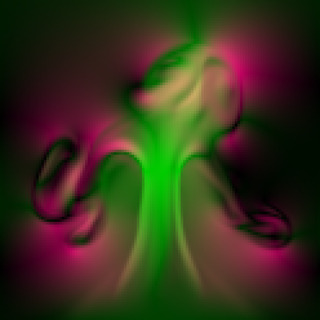}}
  \subcaptionbox*{5$\times$}   {\includegraphics[width=0.24\linewidth]{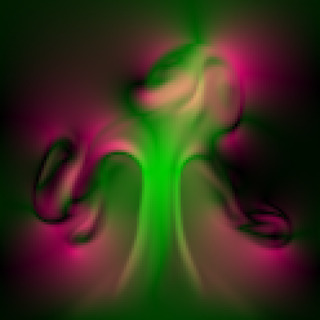}}
  \subcaptionbox*{10$\times$}  {\includegraphics[width=0.24\linewidth]{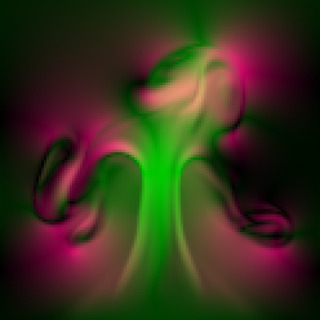}}\\
  \subcaptionbox*{50$\times$}  {\includegraphics[width=0.24\linewidth]{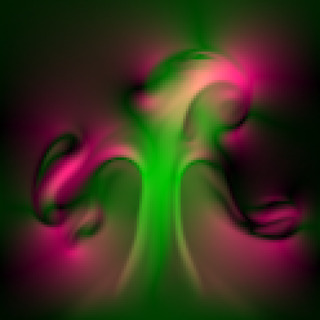}}
  \subcaptionbox*{100$\times$} {\includegraphics[width=0.24\linewidth]{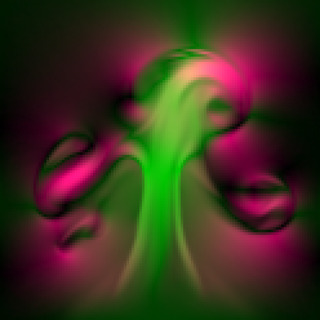}}
  \subcaptionbox*{500$\times$} {\includegraphics[width=0.24\linewidth]{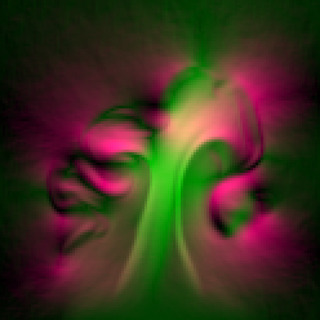}}
  \subcaptionbox*{1000$\times$}{\includegraphics[width=0.24\linewidth]{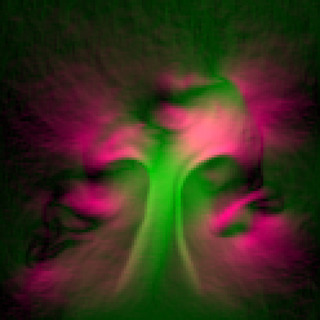}}
  \caption{Rising hot plume: This test compares simulations of a hot plume of
    gas where each simulation adds different levels of noise with the multiplier
    shown below each image. The reference data set is the noise-free result, and
    larger amounts of noise lead to stronger non-linear deviations from the
    reference. The images visualize the x and y velocity components in purple
    and green, respectively.}
  \label{fig:plume}
\end{figure}

As a second setup, a rising plume setup is simulated with a two-dimensional
Eulerian \ac{simple} solver \mycite{patankar1983calculation} using a Boussinesq
approximation. It employs a marker-and-cell grid \mycite{harlow1965numerical} in
conjunction with the fractional step method \mycite{kim1985fract} and Chorin
projection \mycite{chorin1968numerical}. The computational grid has a resolution
of $128^2$ quadratic cells, and the variance shown in the visualizations of
\myfigref{fig:plume} is generated by adding differently scaled white noise to
both components of the velocity. The lowest amount for the version denoted by
1$\times$ uses random values in the range of [0, 10$^{-4}$]. A buoyancy force of
$10^{-3}$ per timestep is employed, and, due to the transient nature of the
simulation, a snapshot after 100 timesteps is visualized. In the visualizations,
noise levels of 5$\times$ and 1$\times$ lead to barely visible changes, which,
nonetheless, influenced the mean assessments of participants in the
corresponding user study.

\subsection{Airfoil}
Lastly, we simulate a steady-state flow around an airfoil. This data set is
likewise simulated with \iac{simple} solver \mycite{patankar1983calculation},
but, in this case, an irregular triangle mesh with 42,730 nodes and 126,964
elements was used for the computational grid. The simulation is solved for
Reynolds-average steady-state and employs the Spalart-Allmaras one equation
turbulence model \mycite{spalart1992one} with a kinematic viscosity $\nu$ of
10$^{-5}$. Angles of attack vary for a freestream velocity magnitude
$|\vec{v}_{\text{free}}|$ of 85, yielding an average Reynolds number $Re$ of
4.25 million across the simulated variants. The airfoil uses the profile
\texttt{GOE-633} obtained from the UIUC database \mycite{uiucDb}.

The visualization for pressure uses the perceptually linear color profile
\emph{``magma''}, while the two velocity components intentionally use a regular
color profile \emph{``jet''}. Our results show that the latter does not pose
problems for our \ac{hvs}-based evaluation.

\section{Results and discussion}

We establish the effectiveness of the \ac{hvs} as a means of accuracy evaluation
for numerical simulations through a series of user studies. Our user studies
employ the \ac{2afc} methodology outlined above, in which participants are asked
which of two candidates they consider to be closer to a third candidate shown as
reference. Each of these candidates typically consists of field data obtained
with one of the numerical methods outlined above. For a data set comparing
different candidates, each participant answers all pairwise questions twice in a
user study where the alternatives for each question are shown in random spacial
order. All of our user studies were performed with 50 participants per study,
which we have found to be sufficient for robust assessments with low variance.
Overall, we collected 56,300 answers of 567 different participants from 55
countries via crowdsourcing.

Due to the simplicity of each task, these studies are very easy to set up and
incur only a small cost. In our setup for a study with six evaluation
candidates, the cost was 33,- USD. In addition, our study results were very
robust. Based on the repeated questions, we measured reliability by computing
the matching rate of answers, which was 83\% on average for all studies in
Experiment 1. Moreover, we evaluated three different crowdsourcing platforms and
found that their results for the same data set are very consistent. Here, the
correlations of the results from different platforms were all greater than 0.95
($P<0.001$).

\subsection{Experiment 1: \ac{hvs} as evaluation metric}

\begin{figure*}[tbp]
  \captionsetup[subfigure]{aboveskip=1pt,belowskip=0pt}
  \centering
  \subcaptionbox*{Breaking dam\label{fig:tset:dam}}               {\includegraphics[height=.152\linewidth]{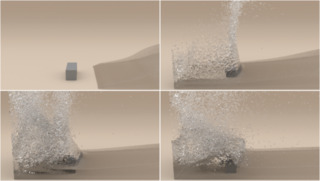}}
  \subcaptionbox*{Plume\label{fig:tset:plume}}                    {\includegraphics[height=.152\linewidth]{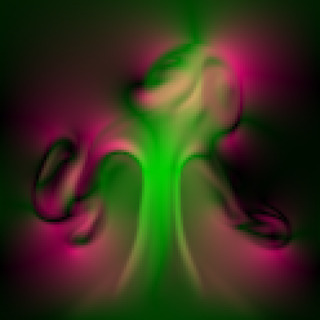}}
  \subcaptionbox*{Airfoil\label{fig:tset:airfoil}}                {\includegraphics[height=.152\linewidth]{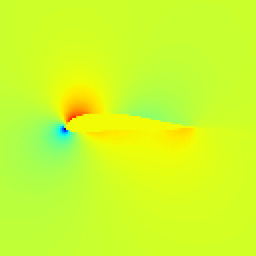}}
  \subcaptionbox*{Viscous shock\label{fig:tset:viscous}}          {\includegraphics[height=.152\linewidth]{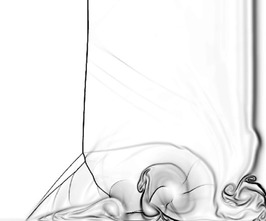}}
  \subcaptionbox*{Double Mach reflection\label{fig:tset:inviscid}}{\includegraphics[height=.152\linewidth]{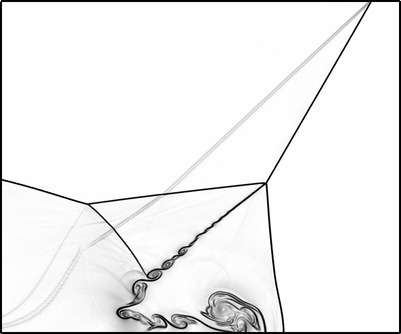}}
  \caption{A selection of fluid simulation evaluations: The images show the
    visualized simulation results where a control parameter such as resolution
    or velocity boundary condition is modified in increments to yield a sequence
    with known ordering. Simulation setups and visualizations for each case
    differ substantially. The full data sets for each example are shown in
    \myfigref{fig:dam} and \ref{fig:plume}--\ref{fig:inviscid-shock-res},
    respectively.}
  \label{fig:tset}
\end{figure*}

\begin{figure*}[tbp]
  \captionsetup[subfigure]{aboveskip=1pt,belowskip=0pt}
  \centering
  \subcaptionbox{\label{fig:tset:graph:dam-flip}Breaking dam with \acs{flip}}{\includegraphics[width=.24\linewidth]{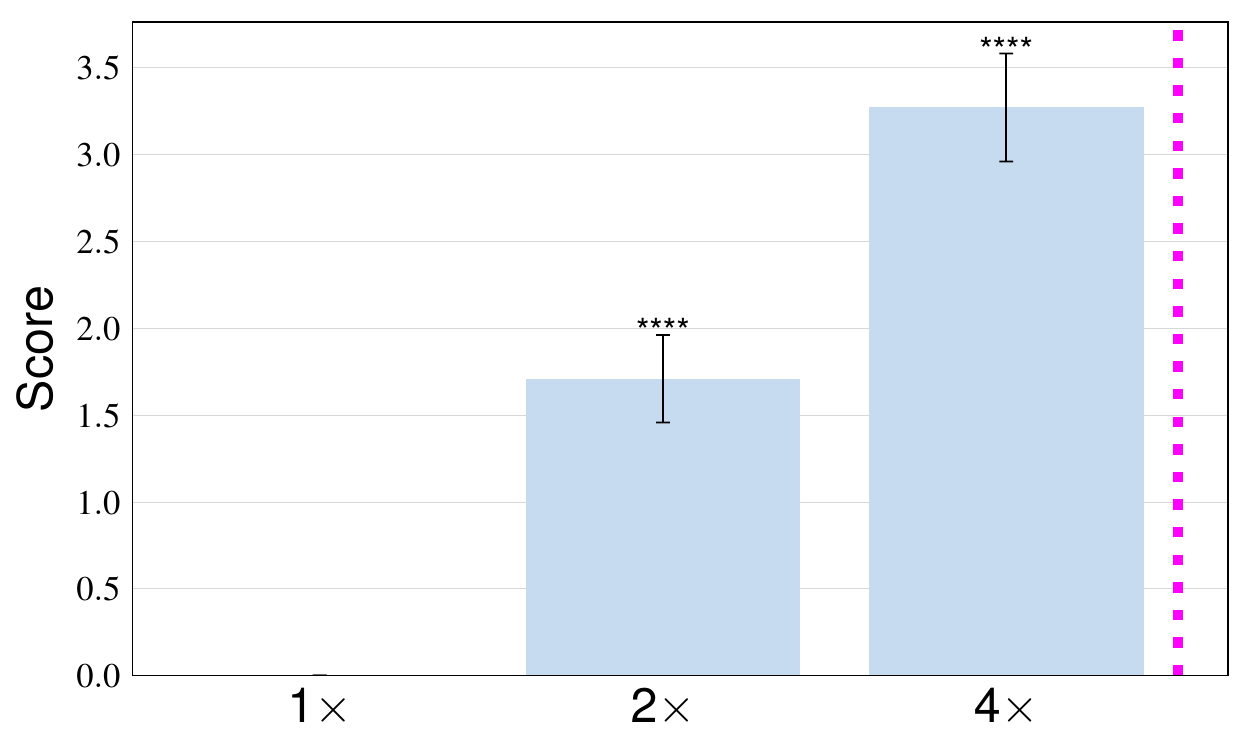}}
  \subcaptionbox{\label{fig:tset:graph:dam-sph}Breaking dam with \acs{sph}}  {\includegraphics[width=.24\linewidth]{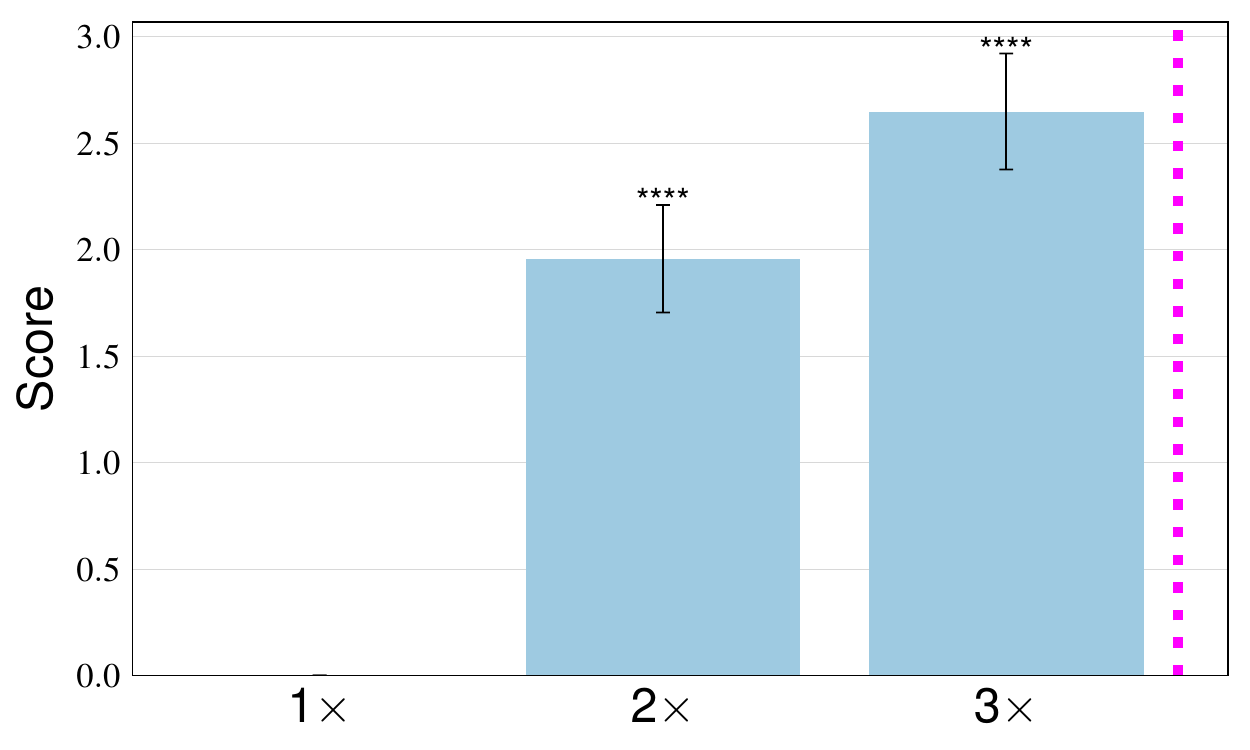}}
  \subcaptionbox{\label{fig:tset:graph:plume}Plume}                    {\includegraphics[width=.24\linewidth]{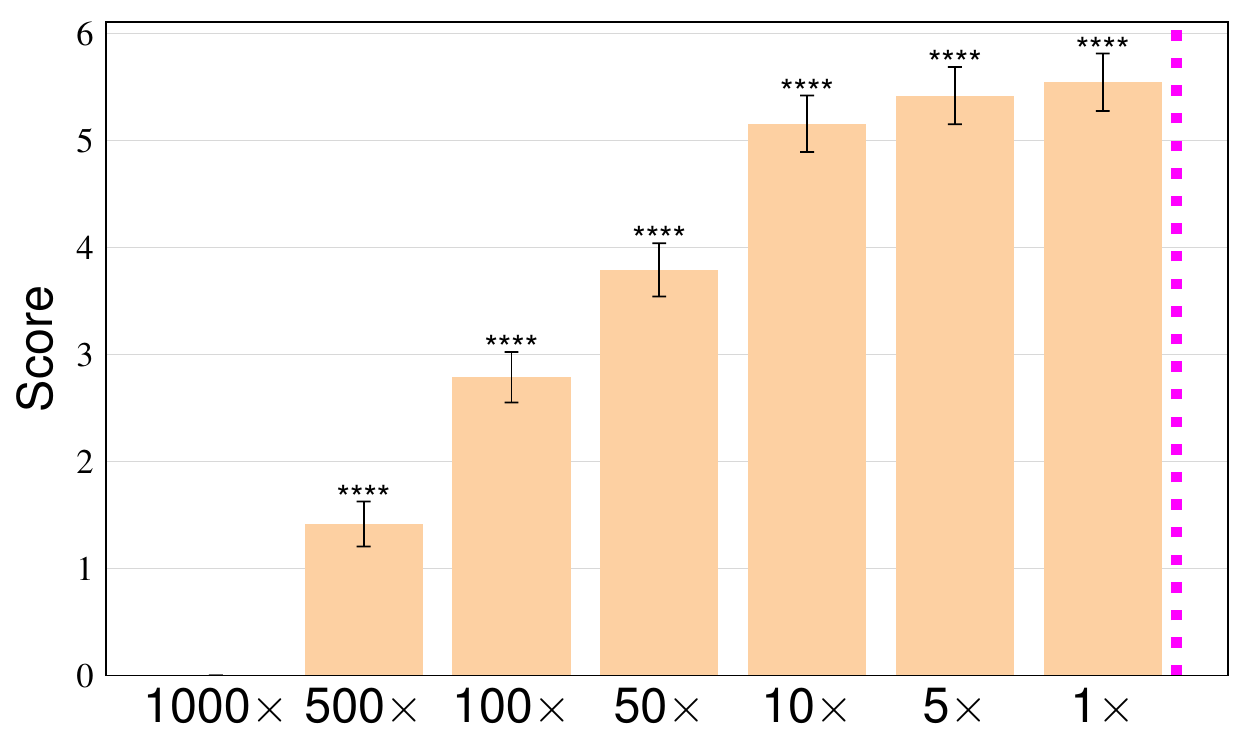}}
  \subcaptionbox{\label{fig:tset:graph:airfoil:p}Airfoil (pressure)}   {\includegraphics[width=.24\linewidth]{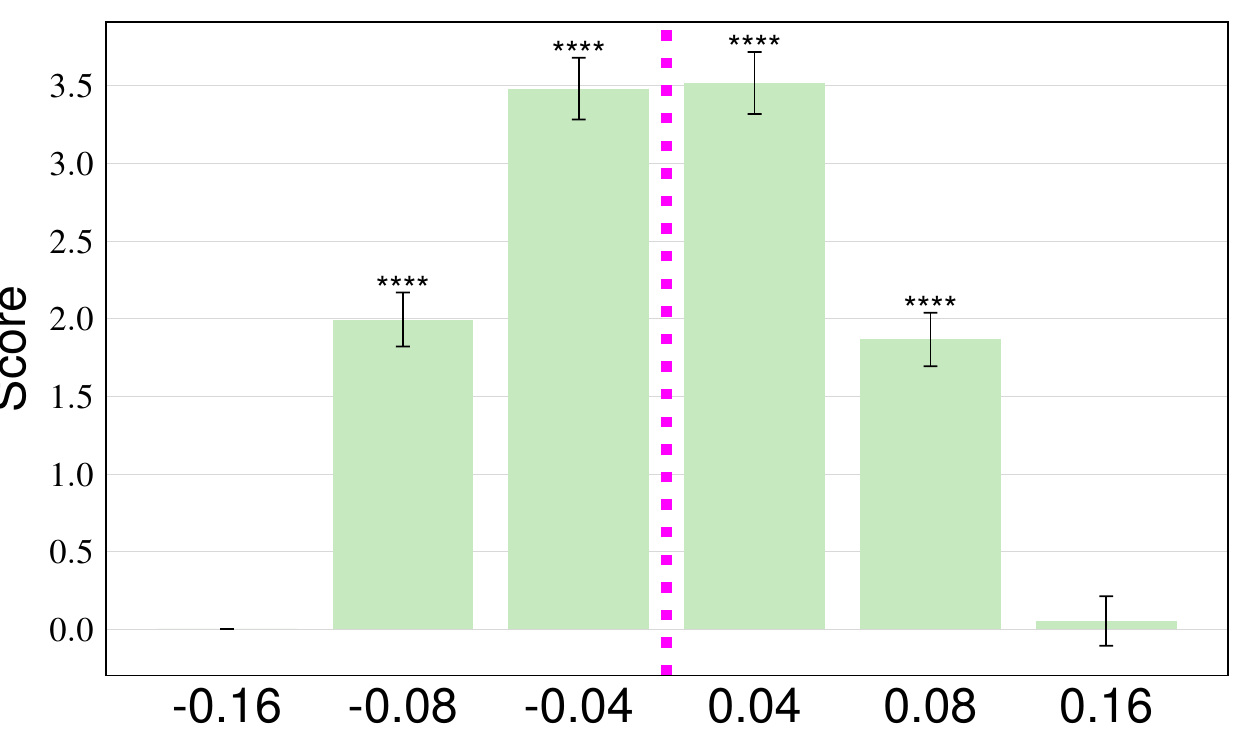}}\\
  \subcaptionbox{\label{fig:tset:graph:airfoil:u}Airfoil (x-velocity)} {\includegraphics[width=.24\linewidth]{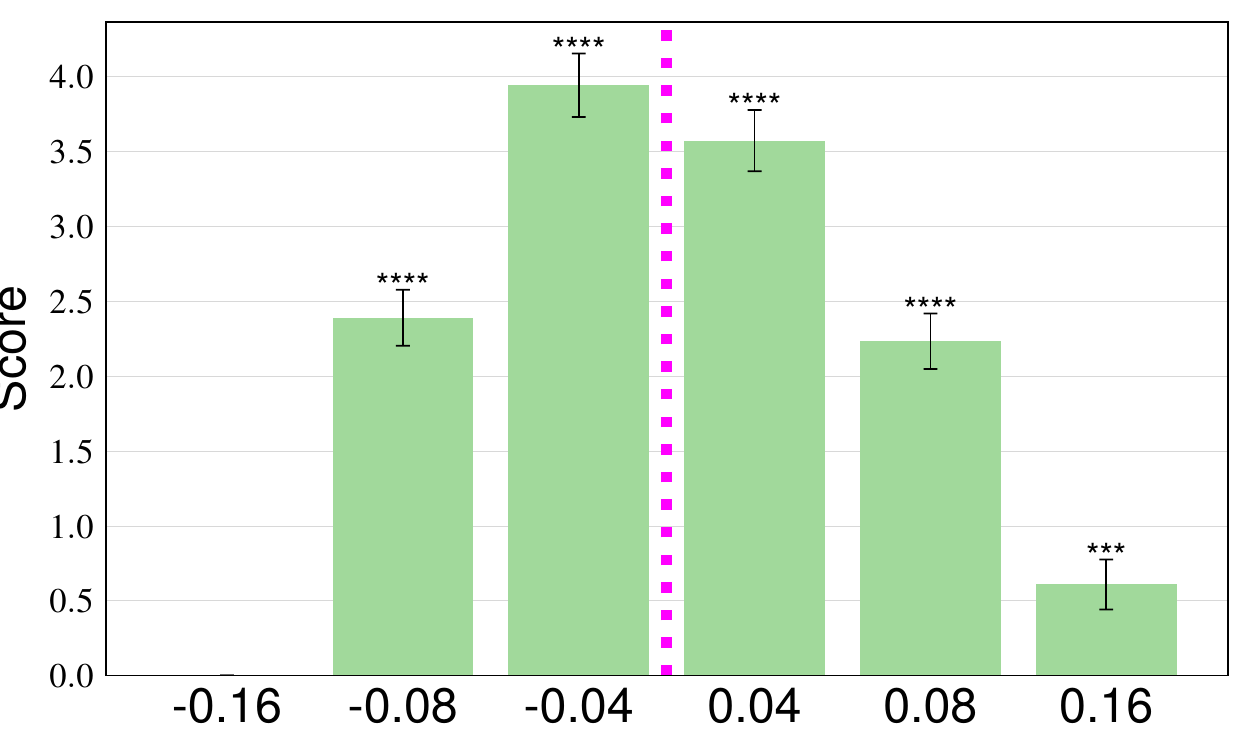}}
  \subcaptionbox{\label{fig:tset:graph:airfoil:v}Airfoil (y-velocity)} {\includegraphics[width=.24\linewidth]{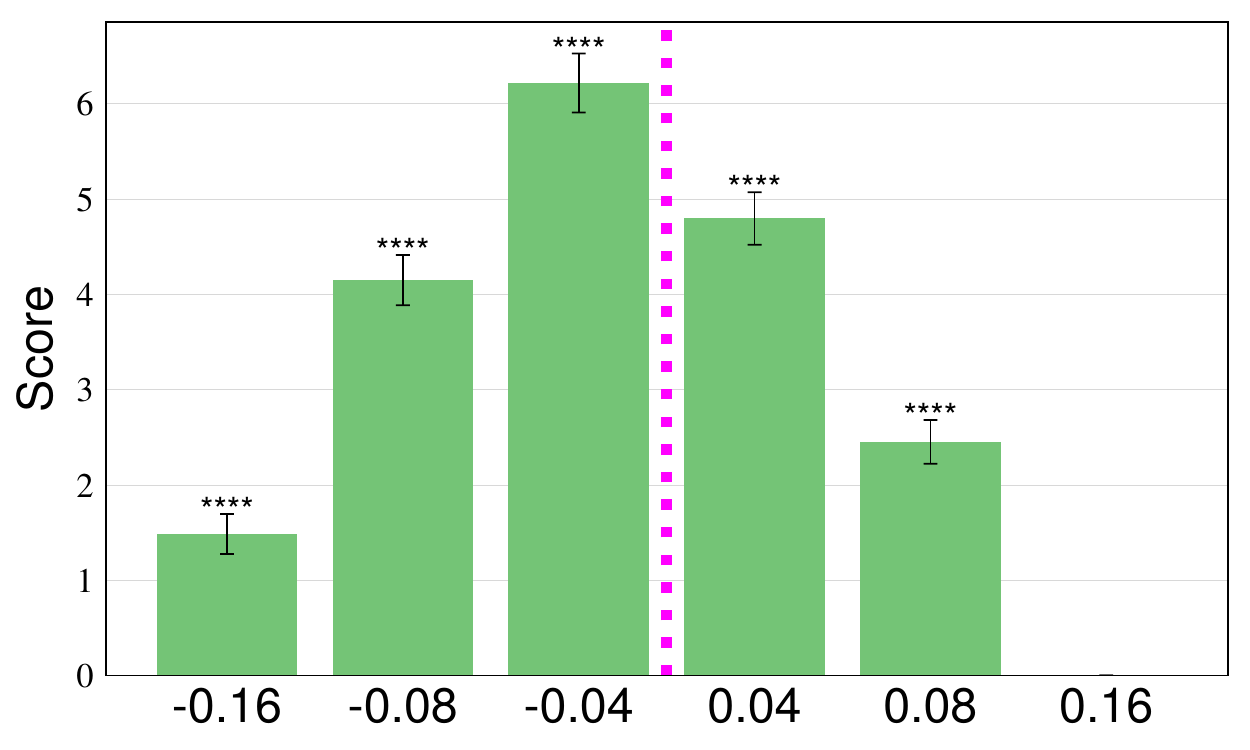}}
  \subcaptionbox{\label{fig:tset:graph:viscous}Viscous shock}          {\includegraphics[width=.24\linewidth]{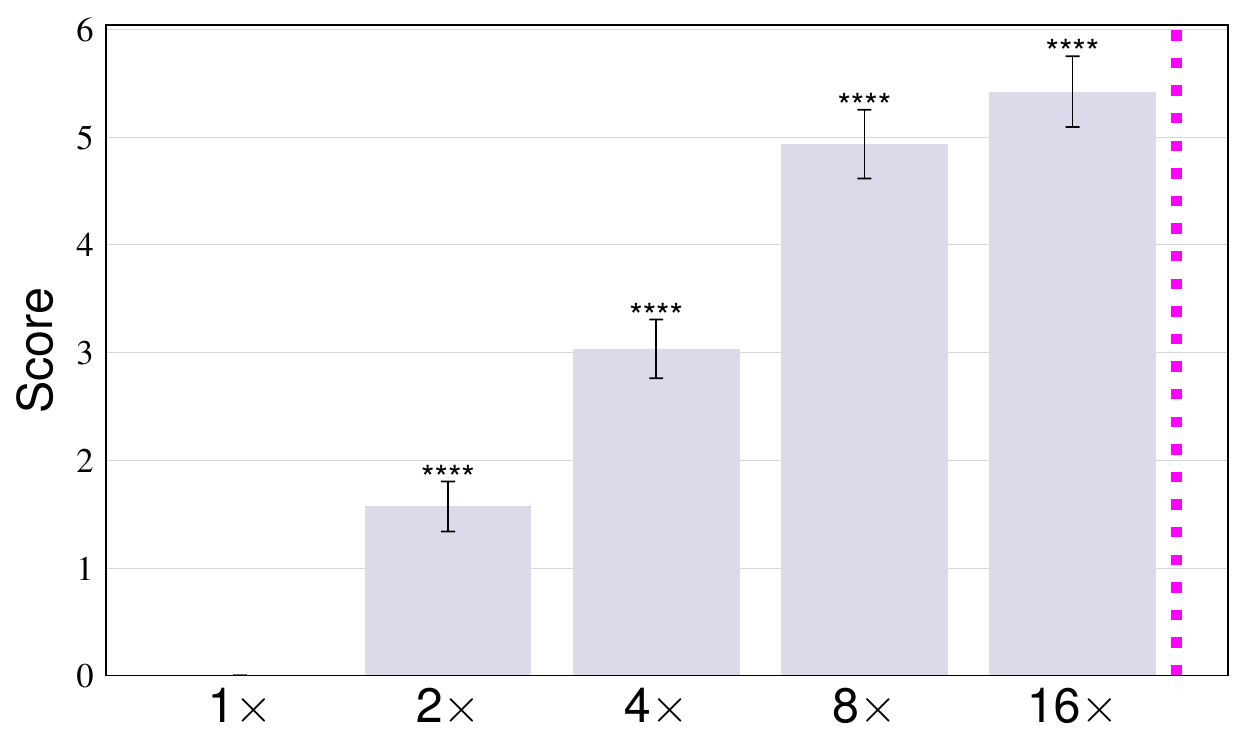}}
  \subcaptionbox{\label{fig:tset:graph:inviscid}Double Mach reflection}{\includegraphics[width=.24\linewidth]{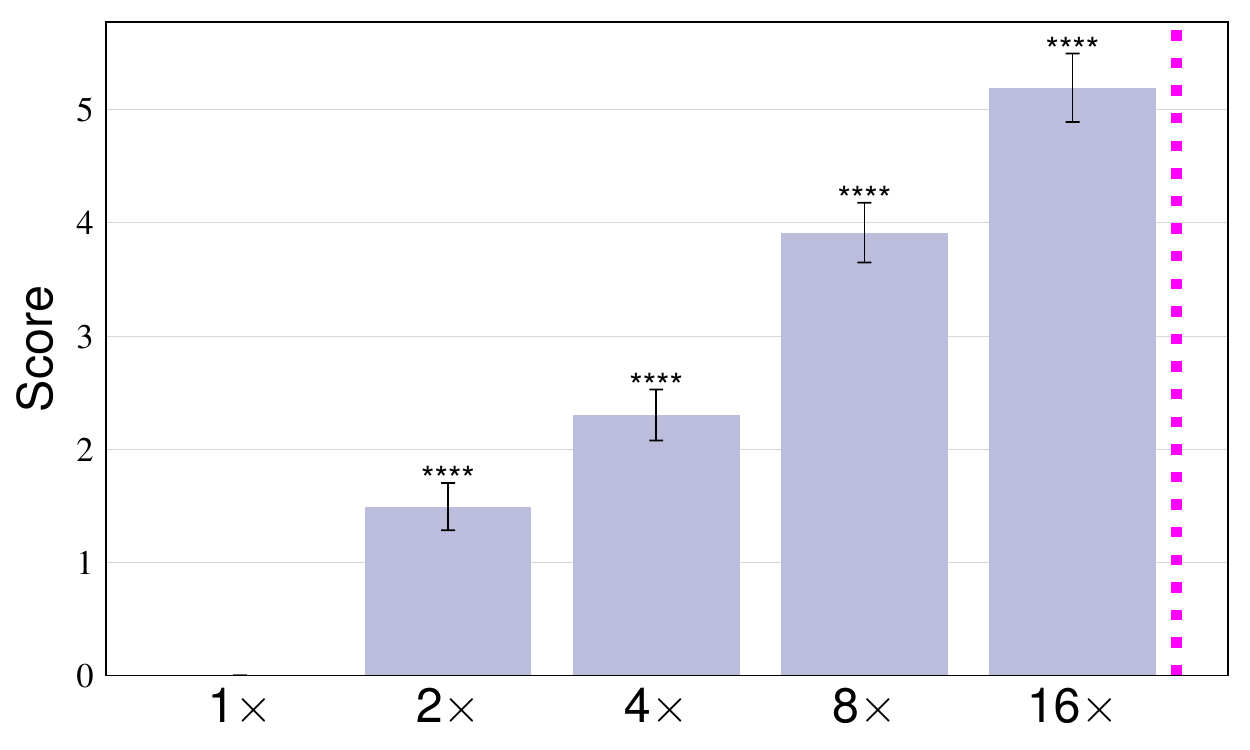}}
  \caption{Accuracy evaluations for the different simulation setups shown in
    \myfigref{fig:tset}: The dashed magenta line indicates the baseline for each
    evaluation. Thus, we anticipate that the performance score of a simulation
    is higher if it is closer to the baseline. The participants of each study
    recover the ground truth ordering without any errors in the established
    ranks. ***$P<0.001$ and ****$P<0.0001$.}
  \label{fig:tset:graphs}
\end{figure*}

It is widely acknowledged that the \ac{hvs} performs extremely well at finding
the similarities and differences between two images
\mycite{cornsweet1970,neri2002}. This property naturally motivates employing it
for quality assessments of typical image processing tasks such as compression
and synthesis \mycite{wang2006}. Our experiments demonstrate that \ac{hvs}-based
evaluations can correctly recover the ground truth ranks for a broad range of
simulation cases and visualization types as shown in \myfigref{fig:tset}. We
have investigated a breaking dam case with a more realistic depiction in
addition to four test cases with more abstract scientific visualizations. These
visualizations cover a wide variety of simulation types and include a rising hot
plume shown in terms of vorticity, a turbulence simulation around an airfoil
visualized with pressure and velocity profiles, and density visualizations of
shock tube simulations. Whereas participants had a good intuition for the
phenomena shown in the breaking dam scene, we expect the large majority of the
participants to have no understanding of the underlying physics for these four
cases \mycite{battaglia2013}. Our results show that all studies reliably recover
the ground truth ordering for all eight scenarios even for cases with abstract
imagery as shown in \myfigref{fig:tset:graphs}. Thus, such crowd-sourced
evaluations of scientific data pose a reliable way to evaluate sets of
scientific results.

Due to its highly non-linear behavior, it is extremely difficult to predict how
a small change of a parameter affects the result of a flow simulation.
Nevertheless, the results will differ more strongly from a baseline result for
larger variations of that parameter. For example, if we simulate a certain setup
using different resolutions considering the highest resolution as the baseline
(i.e., ground truth), the lowest resolution will differ most from the baseline
due to the largest accumulated numerical errors. We conduct a series of user
studies from five different examples where we know the true rank due to
controlled parameter changes. Due to the changed parameters, the calculated
solutions differ in a variety of ways; for example, they contain shifted
velocity structures, varying vortices locations, and different configurations of
transported passive quantities. We chose a representative set of typically
evaluated physical variables, such as density and velocity, which were
visualized either realistically or according to common practice in the
respective scientific community.

\myfigref{fig:tset:graphs} summarizes the resulting ranks of the visualized data
sets with respect to their similarity to a chosen ground truth data set. Here,
the dashed magenta line indicates the location of this ground truth in each
parameter space. Thus, the distance to the dashed line directly indicates the
similarity of the solutions, and the monotonically increasing scores towards the
magenta lines show that all studies recover the expected ordering. As a
consequence, all studies yield a Kendall rank correlation coefficient
\mycite{kendall1983} of one against the ground truth ranks.

\begin{figure}
  \captionsetup[subfigure]{aboveskip=1pt,belowskip=3pt}
  \centering
  \subcaptionbox*{Specifications [\si{m}]}{\includegraphics[width=0.49\linewidth]{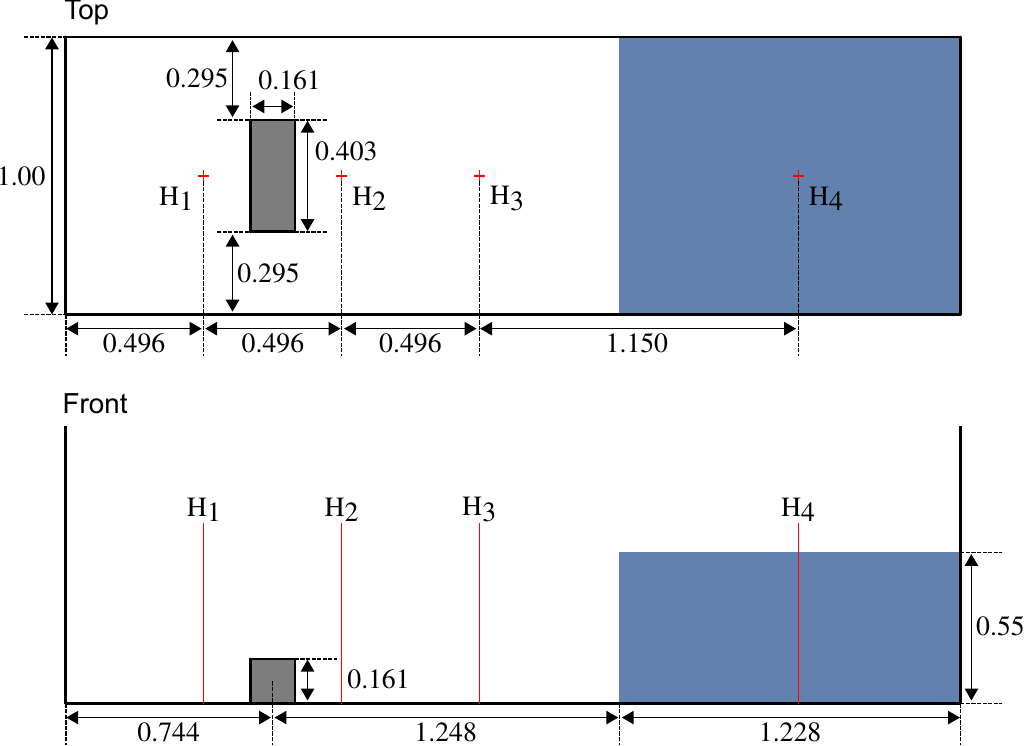}}
  \subcaptionbox*{Reference}              {\includegraphics[width=0.49\linewidth]{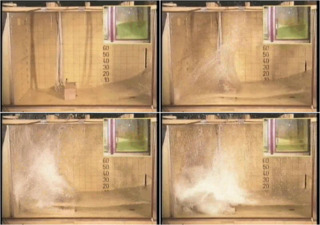}}\\
  \subcaptionbox*{\acs{flip} (1$\times$)} {\includegraphics[width=0.325\linewidth]{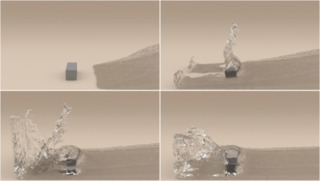}}
  \subcaptionbox*{\acs{flip} (2$\times$)} {\includegraphics[width=0.325\linewidth]{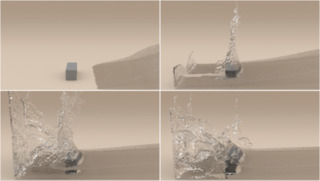}}
  \subcaptionbox*{\acs{flip} (4$\times$)} {\includegraphics[width=0.325\linewidth]{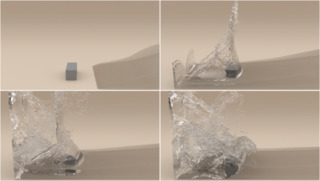}}\\
  \subcaptionbox*{\acs{sph} (1$\times$)}  {\includegraphics[width=0.325\linewidth]{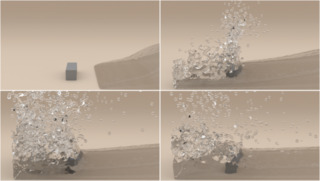}}
  \subcaptionbox*{\acs{sph} (2$\times$)}  {\includegraphics[width=0.325\linewidth]{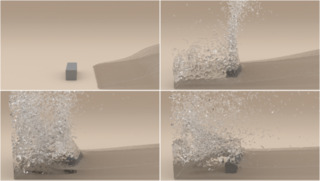}}
  \subcaptionbox*{\acs{sph} (3$\times$)}  {\includegraphics[width=0.325\linewidth]{figs/exp.img4/transA-wbcsph-075}}
  \caption{Breaking dam: The top left shows the specifications of this
    experimental setup \mycite{kleefsman2005}. The probe positions for graphs in
    \myfigref{fig:dam:measure} are denoted as H$_{1\textrm{-}4}$. The reference
    image (top right) shows four selected frames of a video recorded from the
    real-world experiment \mycite{spheric}. The images in the bottom two rows show
    the corresponding frames of simulations from two methods (i.e., \acs{flip} and
    \acs{sph}) with different resolutions. (The resolution multiplier is shown below
    each image.)}
  \label{fig:dam}
\end{figure}

As a first real-world validation case, we consider a \emph{breaking dam} setup,
which contains a large scale wave and splash forming in a rectangular container.
The breaking dam represents a widely adopted simulation case for free surface
flow simulations \mycite{spheric}. We simulate this case with two representative
methods, i.e., \iac{sph} \mycite{adami2012} and \iac{flip} \mycite{zhu2005}
solvers, using three different resolutions per method. For this simulation, we
compare four frames from a realistic visualization of the water surface as shown
in \myfigref{fig:dam} with the corresponding video frames of the real-world
experiment.

Traditionally, the accuracy of a simulation for this breaking dam setup
evaluated with four point-wise probe measurements. However, this data is highly
inconclusive for the simulation data under consideration. The \ac{rmse} between
the experiment and simulation data is shown in Table~\ref{tab:dam:rmse}. This
result is even more pronounced when considering the experimental video and the
simulation visualizations for comparisons. Although improved metrics such as the
\ac{psnr} \mycite{wang2009} and the \ac{ssim} \mycite{wang2004} have been
proposed as specialized tools for comparing images, the substantially different
content of the real-world images and visualizations yields quantified
differences that are as inconclusive as the \ac{rmse} values above as shown in
Table~\ref{tab:dam:rmse}. In contrast, our \ac{hvs}-based evaluation methodology
is able to successfully recover the resolution-based ordering of this breaking
dam case for both simulation methods as shown in
\myfigref{fig:tset:graph:dam-flip} and \myfigref{fig:tset:graph:dam-sph}.

\begin{table}
  \centering
  \begin{tabular}{ccccccc}
    \toprule
    Method                 & H$_1$         & H$_2$         & H$_3$         & H$_4$         & \acs{psnr}     & \acs{ssim}    \\
    \midrule
    \acs{flip} (1$\times$) & \maxhi{0.091} & \maxhi{0.036} & 0.042         & 0.044         & \maxhi{14.179} & \maxhi{0.617} \\
    \acs{flip} (2$\times$) & 0.098         & 0.044         & 0.031         & 0.028         & 14.100         & 0.606         \\
    \acs{flip} (4$\times$) & 0.108         & 0.045         & \maxhi{0.020} & \maxhi{0.026} & 14.024         & 0.590         \\
    \midrule
    \acs{sph} (1$\times$)  & 0.186         & \maxhi{0.024} & \maxhi{0.023} & 0.023         & \maxhi{14.021} & \maxhi{0.581} \\
    \acs{sph} (2$\times$)  & \maxhi{0.129} & 0.032         & 0.025         & 0.026         & 13.994         & 0.574         \\
    \acs{sph} (3$\times$)  & 0.168         & 0.043         & 0.025         & \maxhi{0.021} & 13.860         & 0.570         \\
    \bottomrule
  \end{tabular}
  \caption{\acs{rmse}, \acs{psnr}, and \acs{ssim} of breaking dam simulations:
    The \acs{rmse} is computed using the experimental data and corresponding
    simulation values at four probe positions, H$_{1-4}$. The \acs{psnr} and
    \acs{ssim} are computed between the real-world image and visualization of
    each simulation. Underlined bold numbers indicate the best performance for
    each of the two methods per column.}
  \label{tab:dam:rmse}
\end{table}

Beyond the breaking dam case, we also examine a variety of simulations where the
target images are more abstract scientific visualizations. The four additional
simulation cases in \myfigref{fig:tset} include a rising hot plume shown in
terms of vorticity, pressure and velocity profiles around an airfoil from a
Reynolds-averaged Navier-Stokes turbulence simulation, and density gradient
magnitudes for viscous shock tube and double Mach reflection setups shown in
grayscale, respectively. Whereas participants will have had a good intuition for
the phenomena shown in the breaking dam scene, we expect the large majority of
the participants to have no understanding of the underlying physics for these
four cases \mycite{battaglia2013}.

For the rising plume setup of \myfigref{fig:tset}, we consider different levels
of synthetic noise added to the fluid velocity fields. This represents a
controlled form of artificial error mimicking effects of numerical viscosity.
The ground truth rank with respect to the level of introduced noise is recovered
by participants, and it is noticeable that the curve flattens for noise levels
of 5$\times$ and 1$\times$. These noise levels lead to very slight changes in
the simulated fluid motion and are difficult to discern. Despite being barely
noticeable, the participants, by sake of large numbers, recovered the right
ordering of the different simulation results.

The studies of \myfigsref{fig:tset:graph:airfoil:p}{fig:tset:graph:airfoil:v}
show results for user study evaluations performed with different variables of a
classic airfoil flow simulation. Here, the target ranks are established from
different angles of attack with a zero-degree centerline being used as
reference. Hence, the dashed magenta lines in these graphs are located in the
middle, and the graphs show monotonically increasing scores towards this
centerline. The two shock studies in \myfigref{fig:tset} vary simulation
resolutions for a viscous and an inviscid shock dynamics problems. The latter
one is especially interesting as it represents an inviscid flow problem that
does not have numerical convergence in the traditional sense. In both cases, our
\ac{hvs}-based evaluations recover the ordering with respect to simulation
resolution.

\subsection{Experiment 2: Discretization schemes and features}

\begin{figure}[tbp]
  \centering
  \includegraphics[height=.17\linewidth]{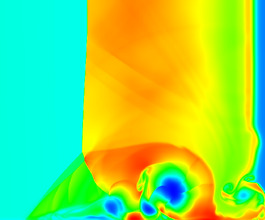}
  \includegraphics[height=.17\linewidth]{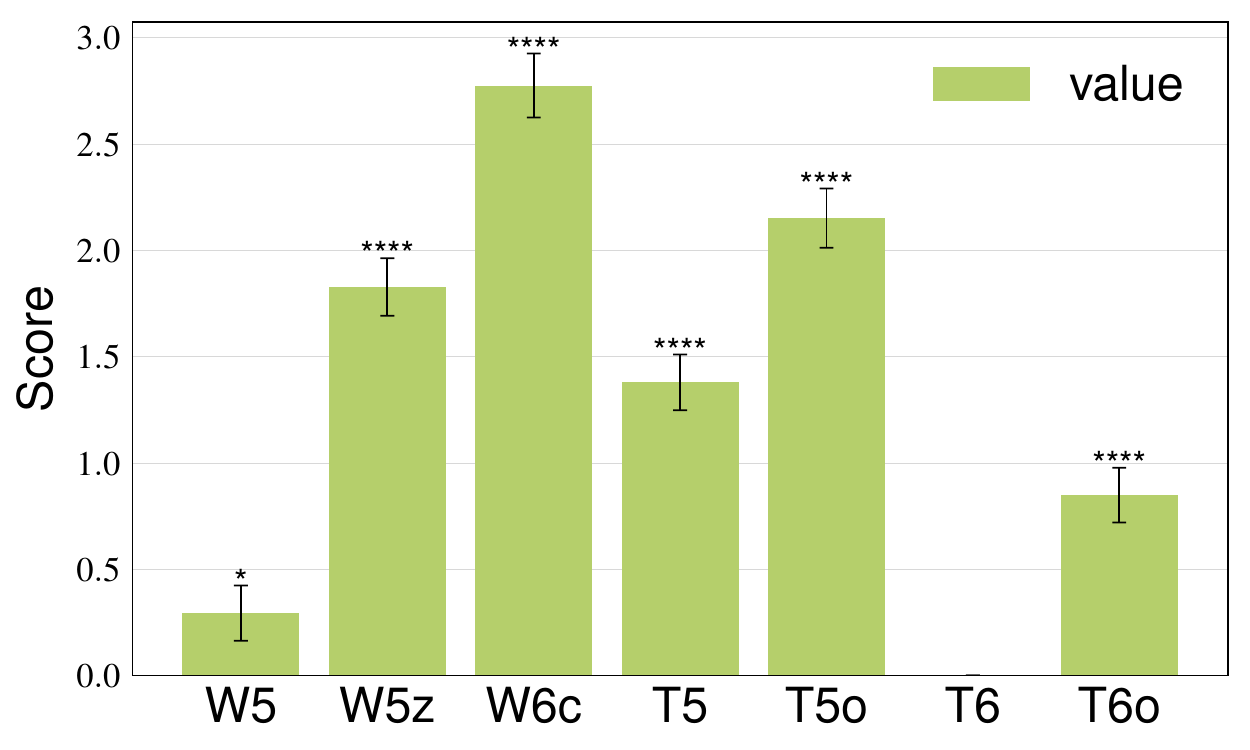}\hfill
  \includegraphics[height=.17\linewidth]{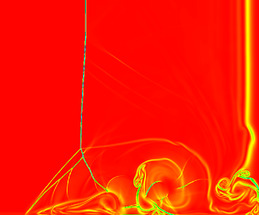}
  \includegraphics[height=.17\linewidth]{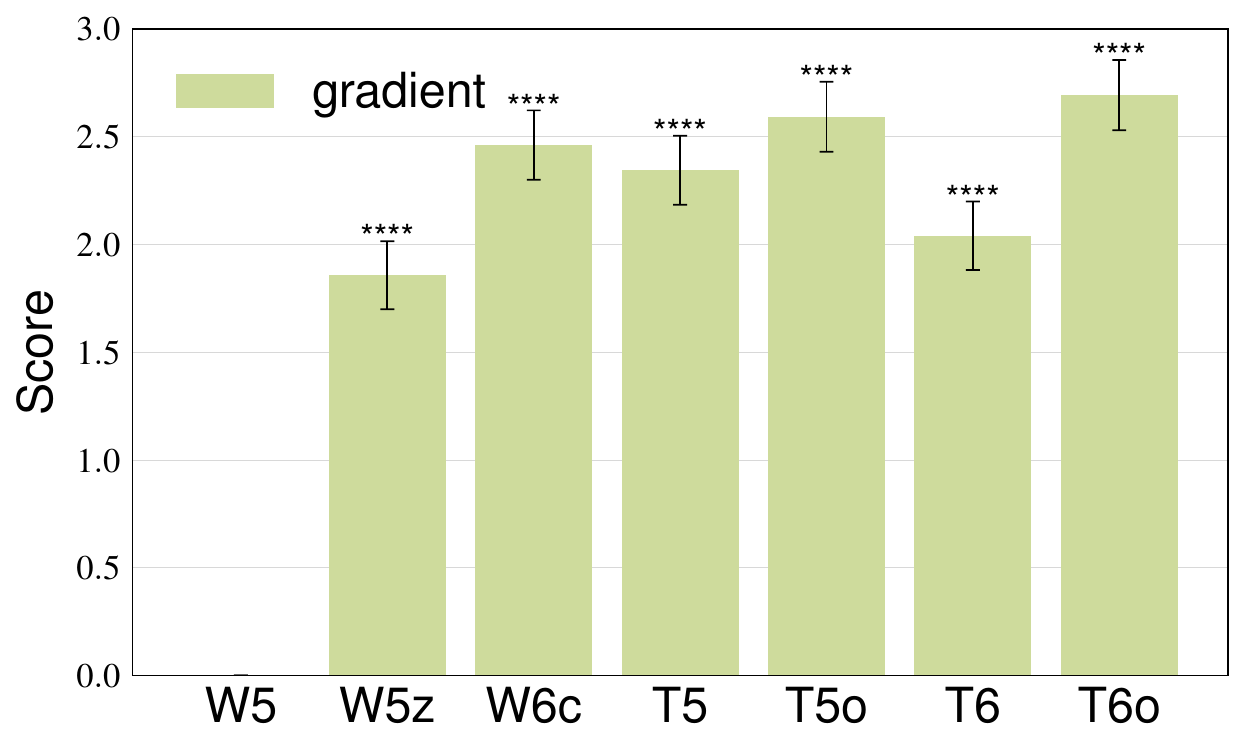}
  \caption{Visualizations of a viscous shock tube simulation and performance
    evaluations of seven finite difference schemes for each visualization: The
    left side shows absolute density values, while the right side visualizes the
    density gradient. ****$P<0.0001$ and *$P<0.05$.}
  \label{fig:seven-methods}
\end{figure}

Having demonstrated the robustness of the \ac{hvs}-based evaluation methodology,
we now turn to analyzing the accuracy of a set of numerical discretization
schemes in the context of two important test cases. More specifically, we
consider seven representative methods of finite difference discretizations from
the class of \ac{eno} schemes. These schemes are very widely used and form the
basis of a vast number of methods for partial differential equation solvers. As
such, comparisons of these methods, in order to establish which methods to
choose for certain applications, have received a significant amount of attention
but have often remained inconclusive \mycite{brehm2015,zhao2014}. In contrast to
these studies, our crowd-sourced evaluations manage to uncover clear differences
between the methods and yield insights that some of the newer schemes under
consideration do not give a stable performance. Our studies illustrate how the
user study methodology can be leveraged to measure the accuracy of simulation
methods (\myfigref{fig:seven-methods}). For these \ac{eno} schemes, it yields an
objective and robust evaluation for a class of methods that have been discussed
controversially in the scientific community.

Here, we present our evaluation results for seven representative methods of
finite difference discretizations from the class of \ac{eno} schemes. We focus
on methods of fifth and sixth order as these represent the most commonly used
compromise between performance and accuracy. In our study, we included three
\emph{weighted} \ac{eno} schemes: W5, W5z, and W6c. Although the W5 scheme does
not represent the latest state of the art, it was included as a traditional
variant for comparison. In addition, we included four more recent modified
schemes, so-called \emph{targeted} \ac{eno} schemes, denoted as T5, T5o, T6, and
T6o. Here, 5 and 6 in each of the abbreviations indicate the order of the
scheme, and the ``z'', ``c'', or ``o'' suffixes indicate different variants.

We evaluate these schemes with respect to their performance in compressible flow
solvers for a viscous shock tube simulation \mycite{daru2000,zhou2018}. This
setup represents a fundamental experiment for engines and turbines and, as such,
has been widely studied in simulations. The central quantity in these
simulations is the fluid density, which is important in terms of its overall
magnitude and with respect to its gradients. While the former highlights the
density values of larger regions, the gradient emphasizes the position of shock
waves and contact discontinuities.

Hence, in order to analyze both aspects, we visualized the data sets simulated
with each of the seven schemes twice: once using absolute values and once using
the density gradient, each time employing the established way to visualize the
corresponding data (\myfigref{fig:seven-methods}). To generate a reference data
set, we used a high-resolution simulation with the traditional W5 scheme. The
ranks for all schemes of both studies are shown in \myfigref{fig:seven-methods}.

When considering the accuracy of the density gradient, all schemes clearly
outperform the W5 baseline. This is not surprising as W5 is the oldest scheme of
the set and many of the other schemes were specifically developed as an
improvement of it. However, interestingly, the higher order T6 scheme performs
less well than its fifth-order counterparts.

Turning to the accuracy for absolute densities shown in
\myfigref{fig:seven-methods} left, the differences in performance are
significantly more diverse than for the density gradients. The W6c scheme
clearly outperforms all other schemes, and the class of targeted schemes stays
behind the two improved weighted variants (i.e., W5z and W6c). The T6 scheme, in
particular, is identified as deviating strongly from the reference solution and
performs even worse than the W5 baseline.

\subsection{Experiment 3: Localization}

We additionally demonstrate that our methodology can be applied recursively in a
divide-and-conquer fashion to deduce which parts of the data were
discriminative. Starting with a user study evaluation for a chosen number of
full data sets, we perform the evaluation on identical smaller regions of each
data set. We can then correlate the ranks for the localized data set with the
ranks for the whole data set. These localized evaluations naturally force
participants to evaluate the data with respect to subsets of the data and, in
this way, can localize differences in the results.

Regarding the evaluations discussed so far, there is an important yet
challenging open question: \emph{Which parts of a solution led to its assessment
  as accurate or inaccurate?} This is especially important when using such
evaluations to guide future developments rather than for making discrete
decisions. In order to improve a method that produced one of the evaluated
solutions, it is important to localize regions that were discriminative.

While, in general, a stochastic subdivision with overlapping regions would be
ideal, we have found that a simple $2\times2$ subdivision yields excellent
results, i.e., our subdivision scheme follows a quadtree structure. To localize
discriminative regions in the data, we can first compute a reliability factor
for the participant's answers. With our user study methodology, a participant
sees every data pair twice. From these duplicate questions, we can compute a
fraction of matching answers for a given data pair. In the limit of random
answers, i.e., 50\% of votes for each of the two alternatives, we can conclude
that participants were not able to distinguish the two corresponding
visualizations.

Let $I$ be an original image, then $I_i$ represents a cutaway patch of $I$ where
the subscript $i$ denotes a single quadrant. Successively appended subscripts
denote the children of a node; for example, $I_{ij}$ represents a cutaway
quadrant $j$ of a larger cutaway patch $I_i$.

\begin{figure}
  \captionsetup[subfigure]{aboveskip=1pt,belowskip=6pt}
  \centering
  \begin{subfigure}[b]{\linewidth}
    \captionsetup[subfigure]{aboveskip=1pt,belowskip=3pt}
    \centering
    \subcaptionbox*{Reference}{\includegraphics[width=0.24\linewidth]{figs/vis.vshock.dens/ref}}
    \subcaptionbox*{W5}    {\includegraphics[width=0.24\linewidth]{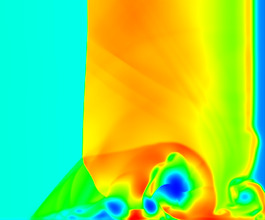}}
    \subcaptionbox*{W5z}   {\includegraphics[width=0.24\linewidth]{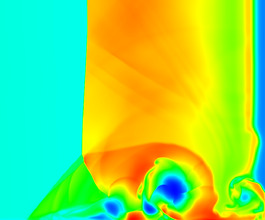}}
    \subcaptionbox*{W6c}   {\includegraphics[width=0.24\linewidth]{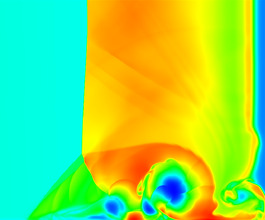}}\\
    \subcaptionbox*{T5}    {\includegraphics[width=0.24\linewidth]{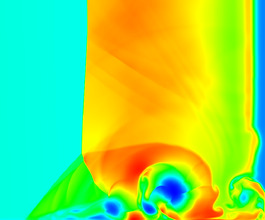}}
    \subcaptionbox*{T5o}   {\includegraphics[width=0.24\linewidth]{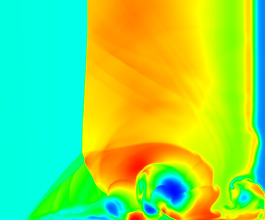}}
    \subcaptionbox*{T6}    {\includegraphics[width=0.24\linewidth]{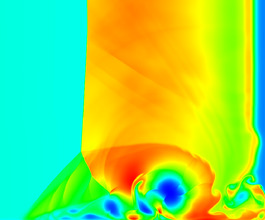}}
    \subcaptionbox*{T6o}   {\includegraphics[width=0.24\linewidth]{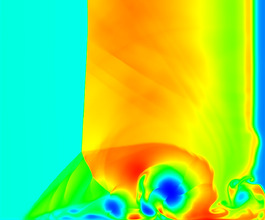}}
    \caption{Density value}
    \label{fig:vis-seven-methods:dv}
  \end{subfigure}
  \begin{subfigure}[b]{\linewidth}
    \captionsetup[subfigure]{aboveskip=1pt,belowskip=3pt}
    \centering
    \subcaptionbox*{Reference}{\includegraphics[width=0.24\linewidth]{figs/vis.vshock.colr/ref}}
    \subcaptionbox*{W5}    {\includegraphics[width=0.24\linewidth]{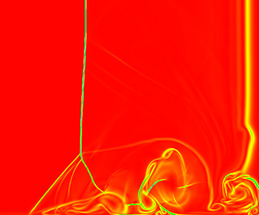}}
    \subcaptionbox*{W5z}   {\includegraphics[width=0.24\linewidth]{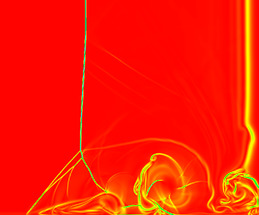}}
    \subcaptionbox*{W6c}   {\includegraphics[width=0.24\linewidth]{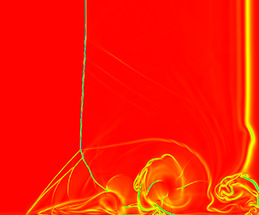}}\\
    \subcaptionbox*{T5}    {\includegraphics[width=0.24\linewidth]{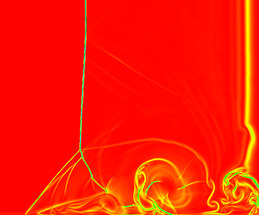}}
    \subcaptionbox*{T5o}   {\includegraphics[width=0.24\linewidth]{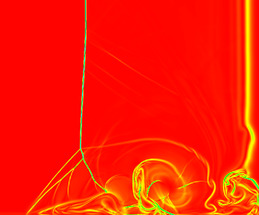}}
    \subcaptionbox*{T6}    {\includegraphics[width=0.24\linewidth]{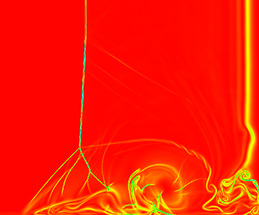}}
    \subcaptionbox*{T6o}   {\includegraphics[width=0.24\linewidth]{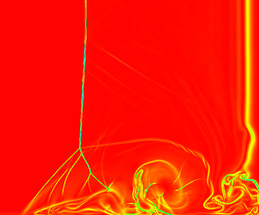}}
    \caption{Density gradient}
    \label{fig:vis-seven-methods:gd}
  \end{subfigure}
  \caption{Viscous shock tubes simulated with different discretization schemes:
    The images (\subref{fig:vis-seven-methods:dv}) and
    (\subref{fig:vis-seven-methods:gd}) show visualizations of the density value
    and the density gradient of each solution, respectively. The grid
    resolutions are 5120$\times$2560 for the reference and 1280$\times$640 for
    the different \acs{eno} schemes.}
  \label{fig:vis-seven-methods}
\end{figure}

We start with four quadrants of the image set $I^\mathrm{dv}$ where the
superscript ``dv'' denotes the density value visualization of
\myfigref{fig:vis-seven-methods:dv}. The set of independent user studies for
each quadrant reveals that two upper quadrants $I_1^\mathrm{dv}$ and
$I_2^\mathrm{dv}$ are less distinguishable. This is intuitively evident as they
contain very few detailed structures. Their average matching rates of answers
were 51.62\% ($\sigma$=7.29\%) and 51.71\% ($\sigma$=9.29\%), respectively. Note
that the matching rates for the two lower quadrants $I_3^\mathrm{dv}$ and
$I_4^\mathrm{dv}$ were 86.38\% ($\sigma$=12.42\%) and 63.71\% ($\sigma$=5.57\%),
respectively. The high matching rate implies that the participants consistently
observe the differences of two patches over the whole set of questions.


On the other hand, if a region results in high reliability and exhibits results
that correlate with the evaluation for the whole data, it strongly indicates
that this region led to the original assessment. For the two lower quadrants of
the viscous shock tube example, we compute the correlation coefficient $\rho$
for their results against the evaluations of the whole data set, i.e., the
original visualization. From the density value studies, we found a strong
positive correlation between the lower left part $I_3^\mathrm{dv}$ and the whole
$I^\mathrm{dv}$, i.e., $\rho(I^\mathrm{dv}, I_3^\mathrm{dv})=0.958$ $(P<0.001)$.
In contrast, the lower right part $I_4^\mathrm{dv}$ is less correlated to the
whole $I^\mathrm{dv}$, i.e., $\rho(I^\mathrm{dv}, I_4^\mathrm{dv})=0.847$
$(P<0.05)$. Drilling down, we also found the strong positive correlations within
the part $I_3^\mathrm{dv}$, i.e.,
$\rho(I^\mathrm{dv}, I_{33}^\mathrm{dv})=0.979$ $(P<0.001)$,
$\rho(I^\mathrm{dv}, I_{34}^\mathrm{dv})=0.98$ $(P<0.001)$, and
$\rho(I_{33}^\mathrm{dv}, I_{34}^\mathrm{dv})=0.996$ $(P<0.00001)$. This
indicates that the part $I_3^\mathrm{dv}$ (and particularly,
$I_{33}^\mathrm{dv} \cup I_{34}^\mathrm{dv}$) is very influential for the
evaluation of the whole simulations.

\begin{figure}[tbp]
  \captionsetup[subfigure]{aboveskip=1pt,belowskip=0pt}
  \centering
  \subcaptionbox*{Target area}{\includegraphics[width=0.24\linewidth]{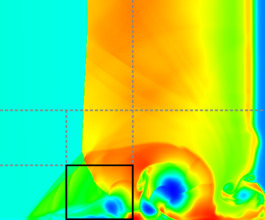}}
  \subcaptionbox*{W6c}        {\includegraphics[width=0.24\linewidth]{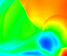}}
  \subcaptionbox*{T6}         {\includegraphics[width=0.24\linewidth]{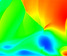}}
  \subcaptionbox*{Reference}  {\includegraphics[width=0.24\linewidth]{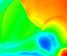}}
  \caption{Localization: For the highlighted target area, the best (W6c) and
    worst (T6) scoring schemes are shown. The T6 solution clearly deviates from
    the reference (shown right).}
  \label{fig:localization:eg}
\end{figure}

Applying our analysis recursively to the lower left quadrant, we can identify
both of its bottom sub-quadrants as discriminative regions. This region
primarily exhibits shock-boundary layer interactions near the bottom wall of the
shock tube. Surprisingly, the key feature that distinguishes the shock tube data
is not the large separation vortex near the bottom center of the visualizations.
Despite this vortex being the most salient feature at first sight, the
participants have identified the lower left ``$\lambda$-shock'' region as a key
difference in the data set, and a post-mortem analysis of the visualizations
confirms this assessment. One of the two sub-quadrants is shown in
\myfigref{fig:localization:eg} for two representative schemes. The performance
evaluation results of this experiment are summarized in
\myfigref{fig:study:cuts}.

\subsection{Experiment 4: Consistency}

One of the most pressing questions for any practitioner in the field of
numerical simulations is to know how reliable and accurate the solution of a
given implementation, method, or solver is. The most common way to study the
behavior of a method is in terms of its convergence order under refinement. It
is usually expressed as the order of the polynomial for the limiting behavior of
the approximation error with respect to the step size. While this expression
reliably indicates the behavior as the step size approaches zero, it hides the
constant factors for the polynomial terms, which are typically unknown and
problem dependent. In practice, it is also often difficult to mathematically
prove the order of convergence for a full algorithm, and, as a consequence, it
is often computed with a series of measurements for a chosen test case. These
measurements again rely on direct vector norms and thus inherently require
smooth and converged solutions \mycite{mehta2016}.

For a solution outside of the convergence regime, i.e., with a finite step size,
the order of convergence does not guarantee that a higher order method actually
performs better than one with lower order. However, with the user study
methodology outlined above, we have a measure of relative accuracy for these
unconverged regimes, and it allows us to establish the notion of a
\emph{\acl{ncm}}, which we will abbreviate as \acs{ncm} in the following. We
define this quantity as the consistency of a method's performance. If a method
repeatedly achieves a high score and outperforms its competitors, this will be
reflected in a small \ac{ncm}. As a consequence, the method also has the highest
chance to perform well when applied to similar new problems. It is worth
pointing out that while our approach is the first one to establish this
consistency in regimes that are not fully converged, just like traditional
tools, it does not guarantee the same evaluation results for new applications.
However, if those applications fall into a similar regime as the one where the
evaluation was performed, there is very high likelihood that the results are
more conclusive than the results of regular convergence studies.

To compute the \ac{ncm}, we employ the \emph{winning probability}. Intuitively,
the winning probability represents the average probability of one method
performing better than the others. The standard deviation of the winning
probability across a series of studies, then, indicates how consistently a
method behaves relative to the others and yields the \ac{ncm}. Thus, the method
with the smallest \ac{ncm} is the one with the lowest standard deviation in
terms of winning probability.

\begin{figure*}[tbp]
  \captionsetup[subfigure]{aboveskip=1pt,belowskip=0pt}
  \centering
  \subcaptionbox*{1$\times$} {\includegraphics[height=.162\linewidth]{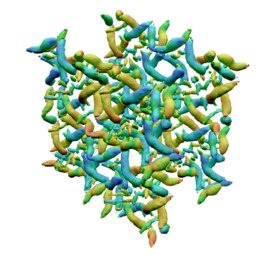}}
  \subcaptionbox*{2$\times$} {\includegraphics[height=.162\linewidth]{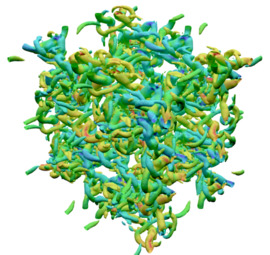}}
  \subcaptionbox*{4$\times$} {\includegraphics[height=.162\linewidth]{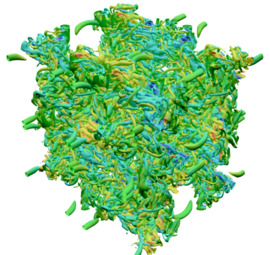}}
  \subcaptionbox*{Reference} {\includegraphics[height=.162\linewidth]{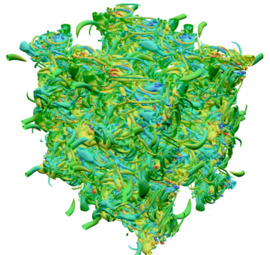}}
  \subcaptionbox*{Mean winning probability}{\includegraphics[height=.162\linewidth]{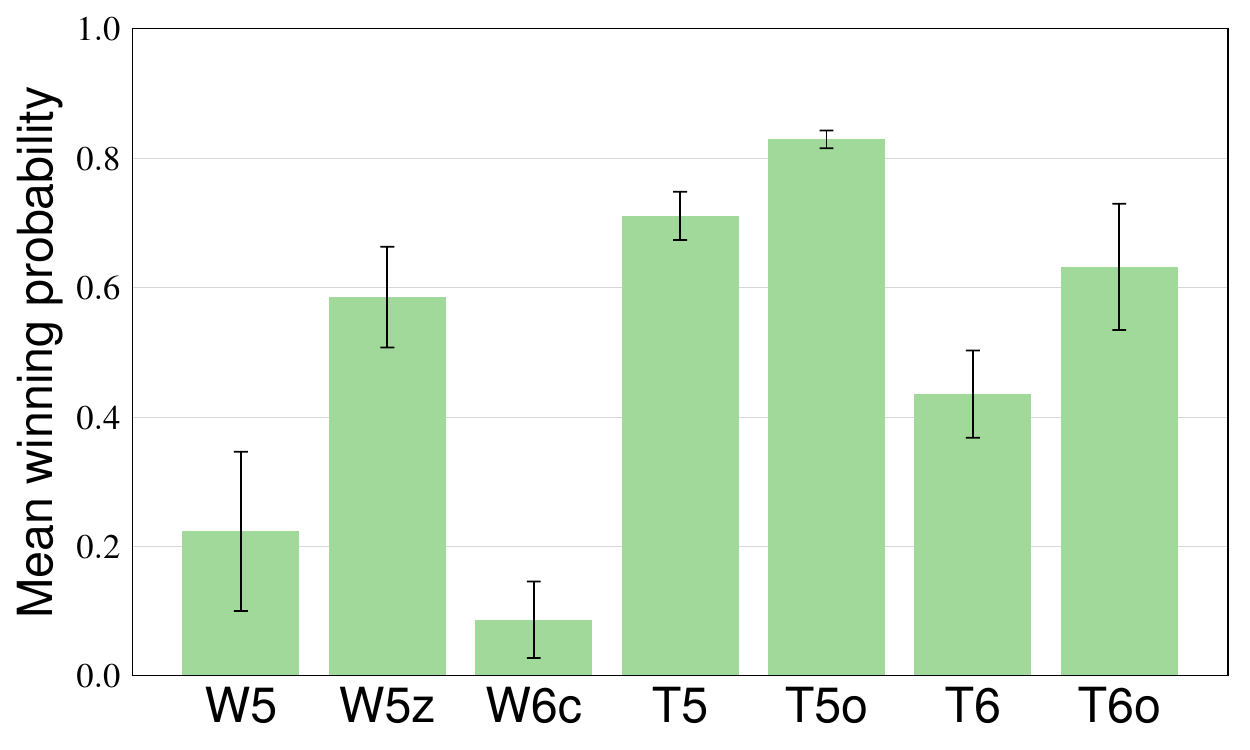}}
  \caption{Different resolutions of the Taylor-Green vortex flow simulation: The
    resolutions are 64$^3$ for 1$\times$, 128$^3$ for 2$\times$, 256$^3$ for
    4$\times$, and 512$^3$ for Reference. The images show the isosurface for a
    Q-criterion value of three and are colored by x vorticity magnitude. The
    graph shows mean and standard deviation of winning probabilities for seven
    schemes across these resolutions.}
  \label{fig:tgv}
\end{figure*}

We have measured the \ac{ncm} for the aforementioned \ac{eno} schemes in two
very different simulation settings: first in the context of incompressible
turbulence and second in the context of compressible shocks. We consider the
incompressible case with three-dimensional Navier-Stokes simulations of
\iac{tgv} problem \mycite{taylorGreen} with a Reynolds number of 3,600
\mycite{brachet1983} that leads to complex decaying vortex structures as shown
in \myfigref{fig:tgv}. The compressible case is studied with a set of
two-dimensional shock tube simulations, similar to those shown in
\myfigref{fig:seven-methods}. Each test case represents an important regime of
fluid flow problems, and our evaluations successfully establish the consistency
of the different simulation methods across a range of different resolutions.

Interestingly, we find that the method that performs worst for the
incompressible case emerges as the winner for the compressible shock flows.
These results do not contradict each other; they rather illustrate how the
complex interplay of choices for numerical solvers can influence the final
results. Obviously, the assessment of this method would have been meaningless in
isolation, and it requires a robust assessment of its accuracy in practical
settings to reveal this behavior. Since both setups represent typical
computational fluid dynamics solvers and resolution ranges, our studies
highlight the importance of accuracy measurements in the near-convergence
regimes, where methods are typically used in practice rather than in the fully
converged limit.

\myfigref{fig:tgv} shows a selection of \ac{tgv} simulation results with
increasing resolution together with the reference solution that was obtained
numerically with a high-resolution simulation. We evaluated each scheme at three
different resolutions, and the corresponding winning probabilities are shown in
the figure together. (See also \mytabref{tab:winp}.) It is immediately apparent
from the graphs that some of the schemes, such as T6o and W5z, exhibit
significant differences across the three resolutions. The T5o scheme, on the
other hand, has a very high and consistent winning probability. Thus, T5o is the
clear winner among the investigated schemes, while the W6c scheme performs worst
despite its high order.

\begin{figure*}[tbp]
  \captionsetup[subfigure]{aboveskip=1pt,belowskip=0pt}
  \centering
  \subcaptionbox*{1$\times$}               {\includegraphics[height=.102\linewidth]{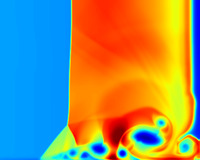}}
  \subcaptionbox*{1.5$\times$}             {\includegraphics[height=.102\linewidth]{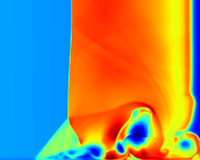}}
  \subcaptionbox*{2$\times$}               {\includegraphics[height=.102\linewidth]{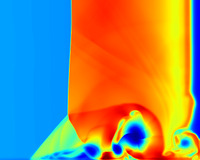}}
  \subcaptionbox*{3$\times$}               {\includegraphics[height=.102\linewidth]{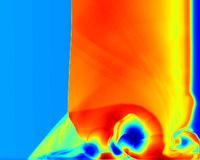}}
  \subcaptionbox*{4$\times$}               {\includegraphics[height=.102\linewidth]{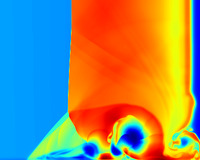}}
  \subcaptionbox*{Reference}               {\includegraphics[height=.102\linewidth]{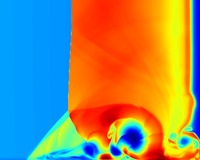}}
  \subcaptionbox*{\scriptsize{Mean winning probability}}{\includegraphics[height=.102\linewidth]{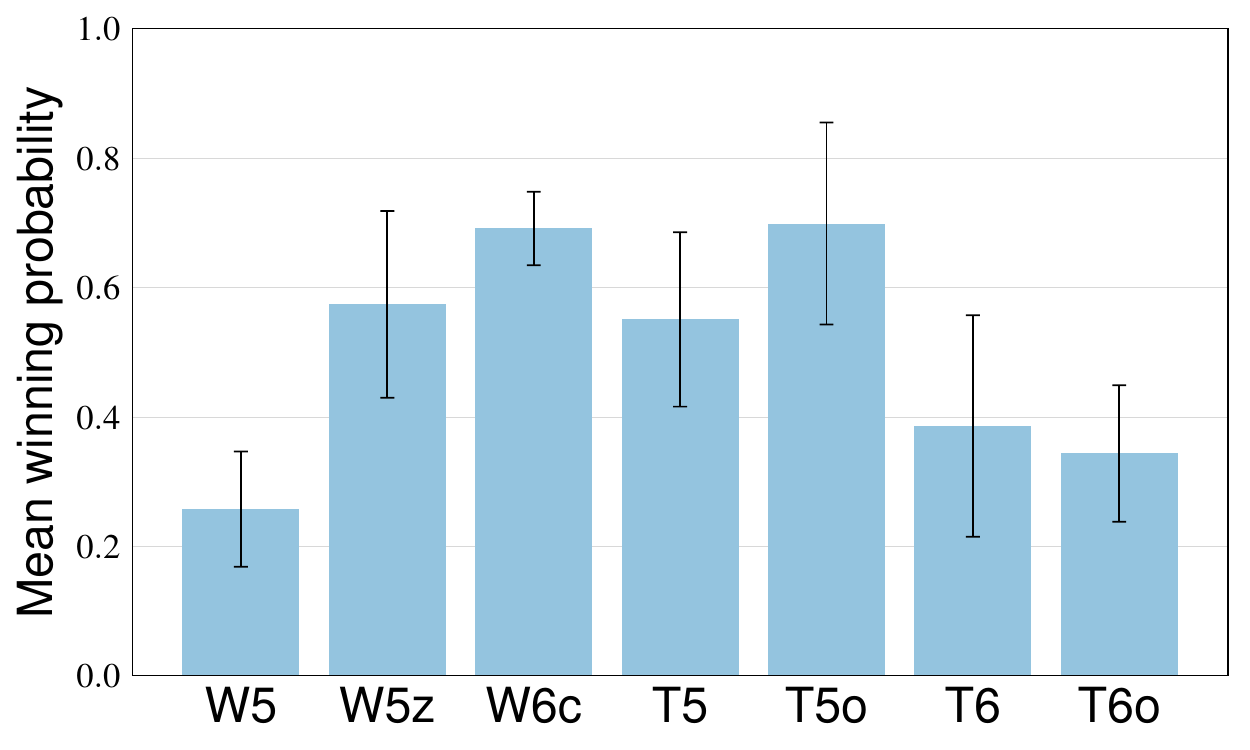}}\\
  \caption{Resolutions used to determine the \ac{ncm} for viscous shock tube
    simulations: From left to right, 1$\times$=640$\times$320,
    1.5$\times$=960$\times$640, 2$\times$=1280$\times$640,
    3$\times$=1920$\times$960, 4$\times$=2560$\times$1280, and Reference is
    5120$\times$2560. The graph shows mean and standard deviation of winning
    probabilities for seven schemes across these resolutions.}
  \label{fig:seven-methods:crossres}
\end{figure*}

We also performed an \ac{ncm} study with an even wider range of resolutions for
the viscous shock tube case from Experiment 2. Here, we used five different
resolutions, which are shown in \myfigref{fig:seven-methods:crossres} together
with their winning probabilities. While T5o has the highest mean of winning
probabilities for these simulations, it also exhibits a very high standard
deviation. Thus, it is very unreliable in terms of its performance for viscous
shock tube simulations. In this case, the W6c scheme has the lowest standard
deviation and thus emerges as the best candidate among the investigated schemes.

\section{Conclusion and outlook}

We have shown that the \ac{hvs} provides a reliable and robust new evaluation
metric that can be employed to analyze a wide range of previously inconclusive
problems. Furthermore, we have shown its usefulness to quantify and measure a
derived property, namely the near-convergence consistency.
  
However, our study only represents a first step towards establishing perceptual
evaluations as an accuracy metric. As such, our approach has limitations that
point towards avenues of future work. For example, for the evaluations of this
study, we have relied on established visualization techniques for the respective
fields our model problems were selected from. This has naturally imposed
knowledge from domain experts. However, it will be highly interesting to further
investigate the influence of visualization on reliability and robustness of our
approach. Moreover, whereas the experiments of this study have focused on fluid
dynamics problems, we envision that this evaluation methodology will be
preferable over simple vector norms for any other problem with complex field
data, such as meteorology and solid mechanics
\mycite{lin1998,cheng2011,zhang2017}.

\begin{table}
  \caption{Winning probability for Taylor-Green vortex flow simulations (top)
    and viscous shock tube simulations (bottom): $\mu$ and $\varepsilon$
    represent the average index and standard deviation (i.e., the \acl{ncm}),
    respectively. The underlined bold number indicates the highest probability
    or lowest error thus the best performance at each column, and the bold
    number indicates the second best one.}
  \label{tab:winp}
  \centering
  \begin{tabular}{cc}
    \begin{tabular}{l|ccc|cc}
      \toprule
      Method & 1$\times$     & 2$\times$     & 4$\times$     & $\mu$         & $\varepsilon$                                 \\
      \midrule
      W5     & .130          & .142          & .397          & .223          & .151                                          \\
      W5z    & .478          & .615          & .662          & .585          & .095                                          \\
      W6c    & .163          & .075          & .020          & .086          & .072                                          \\
      T5     & .710          & .665          & \textbf{.757} & \textbf{.711} & \textbf{.046}                                 \\
      T5o    & \maxhi{.813}  & \maxhi{.847}  & \maxhi{.827}  & \maxhi{.829}  & \maxhi{.017}                                  \\
      T6     & .477          & .488          & .340          & .435          & .082                                          \\
      T6o    & \textbf{.728} & \textbf{.668} & .498          & .632          & .119                                          \\
      \bottomrule
    \end{tabular}
    \vspace{1em}\\
    \begin{tabular}{l|ccccc|cc}
      \toprule
      Method & 1$\times$     & 1.5$\times$   & 2$\times$     & 3$\times$     & 4$\times$     & $\mu$         & $\varepsilon$ \\
      \midrule
      W5     & .127          & .290          & .397          & .262          & .212          & .257          & \textbf{.100} \\
      W5z    & .300          & .638          & \textbf{.645} & .570          & \textbf{.715} & .574          & .161          \\
      W6c    & \textbf{.717} & .675          & .640          & \textbf{.635} & \maxhi{.788}  & \textbf{.691} & \maxhi{.063}  \\
      T5     & .422          & \maxhi{.790}  & .465          & .605          & .470          & .550          & .151          \\
      T5o    & \maxhi{.863}  & \textbf{.750} & \maxhi{.790}  & \maxhi{.680}  & .410          & \maxhi{.699}  & .174          \\
      T6     & .635          & .158          & .235          & .480          & .420          & .386          & .191          \\
      T6o    & .437          & .198          & .328          & .268          & .485          & .343          & .118          \\
      \bottomrule
    \end{tabular}
  \end{tabular}
\end{table}

\section*{Acknowledgments}

This work was supported by the ERC Starting Grant (637014) and National Natural
Science Foundation of China (No. 11628206). We thank Wei Wei of Tsinghua
University for sharing simulation data.

\bibliographystyle{model1-num-names}
\bibliography{references}

\FloatBarrier

\section*{Appendix}

\begin{figure}
  \centering
  \includegraphics[width=.495\linewidth,page=1]{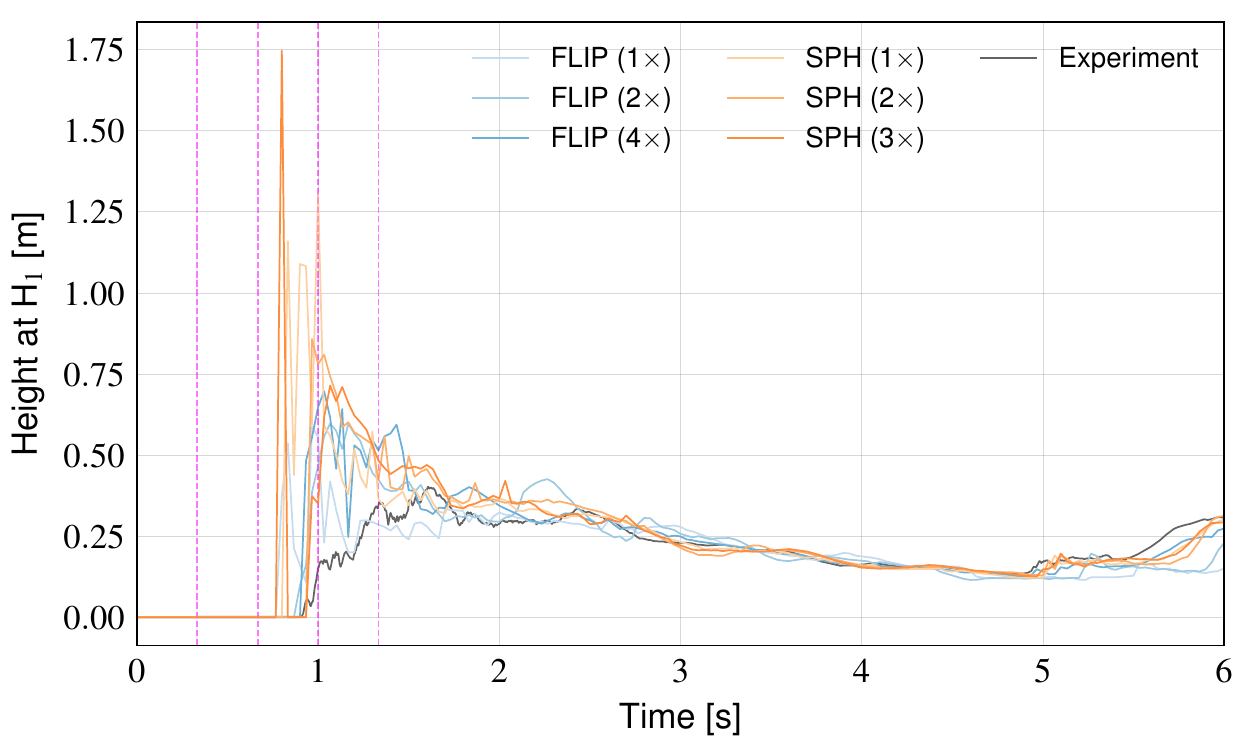}
  \includegraphics[width=.495\linewidth,page=2]{figs/measures}\\
  \includegraphics[width=.495\linewidth,page=3]{figs/measures}
  \includegraphics[width=.495\linewidth,page=4]{figs/measures}
  \caption{Graphs for the breaking dam user study: The graphs plot the water
    height values measured at four probes, H$_{1\textrm{-}4}$, from both the
    experiment (black) and simulations (blue and orange). The dashed vertical
    lines in magenta indicate the selected moments for the images shown in
    \myfigref{fig:dam}. It is difficult to clearly determine which simulation
    performs better (i.e., closer to the black ground truth) from these graphs.}
  \label{fig:dam:measure}
\end{figure}

\clearpage

\begin{figure}
  \centering
  \begin{subfigure}[b]{\linewidth}
    \captionsetup[subfigure]{aboveskip=1pt,belowskip=3pt}
    \centering
    \subcaptionbox*{Reference ($a=0$)}{\includegraphics[width=0.325\linewidth]{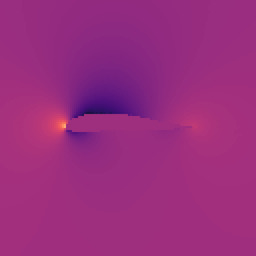}}\\
    \subcaptionbox*{$a=-0.16$}        {\includegraphics[width=0.325\linewidth]{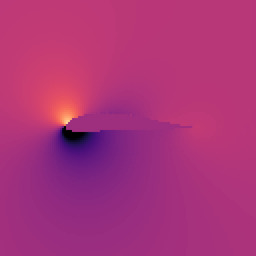}}
    \subcaptionbox*{$a=-0.08$}        {\includegraphics[width=0.325\linewidth]{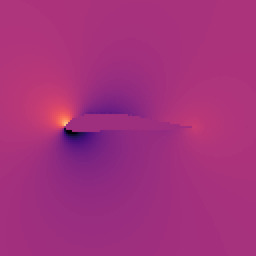}}
    \subcaptionbox*{$a=-0.04$}        {\includegraphics[width=0.325\linewidth]{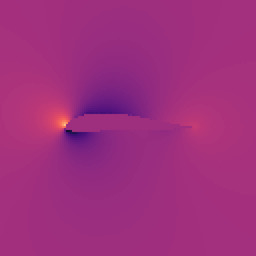}}\\
    \subcaptionbox*{$a=0.04$}         {\includegraphics[width=0.325\linewidth]{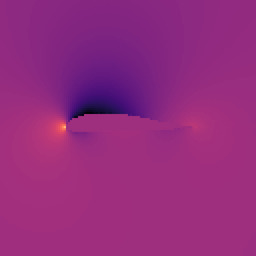}}
    \subcaptionbox*{$a=0.08$}         {\includegraphics[width=0.325\linewidth]{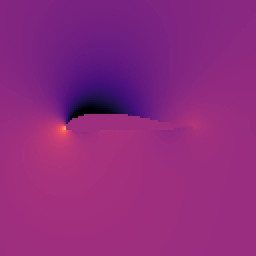}}
    \subcaptionbox*{$a=0.16$}         {\includegraphics[width=0.325\linewidth]{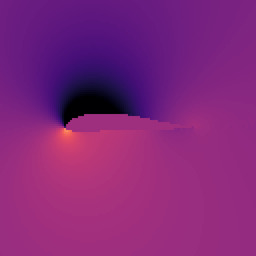}}
    \caption{Pressure}
    \label{fig:airfoil:pres}
  \end{subfigure}
\end{figure}

\begin{figure}\ContinuedFloat
  \centering
  \begin{subfigure}[b]{\linewidth}
    \captionsetup[subfigure]{aboveskip=1pt,belowskip=3pt}
    \centering
    \subcaptionbox*{Reference ($a=0$)}{\includegraphics[width=0.325\linewidth]{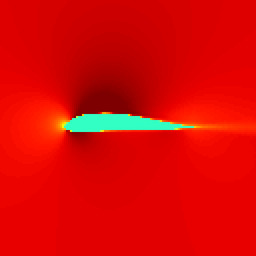}}\\
    \subcaptionbox*{$a=-0.16$}        {\includegraphics[width=0.325\linewidth]{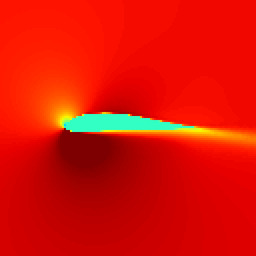}}
    \subcaptionbox*{$a=-0.08$}        {\includegraphics[width=0.325\linewidth]{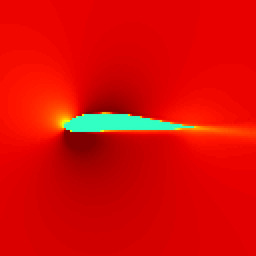}}
    \subcaptionbox*{$a=-0.04$}        {\includegraphics[width=0.325\linewidth]{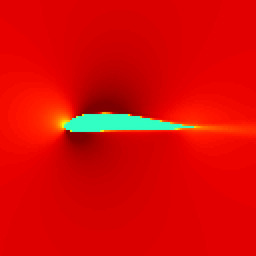}}\\
    \subcaptionbox*{$a=0.04$}         {\includegraphics[width=0.325\linewidth]{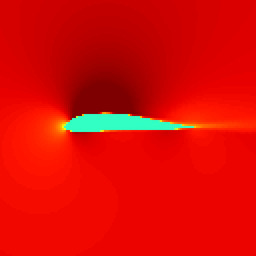}}
    \subcaptionbox*{$a=0.08$}         {\includegraphics[width=0.325\linewidth]{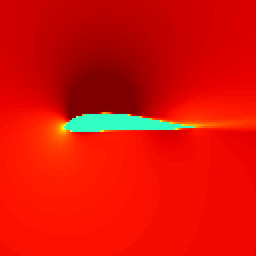}}
    \subcaptionbox*{$a=0.16$}         {\includegraphics[width=0.325\linewidth]{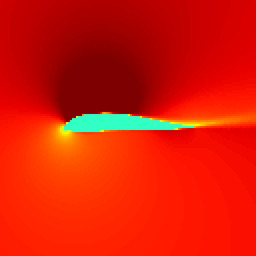}}
    \caption{X-velocity}
    \label{fig:airfoil:velx}
  \end{subfigure}
\end{figure}

\begin{figure}\ContinuedFloat
  \centering
  \begin{subfigure}[b]{\linewidth}
    \captionsetup[subfigure]{aboveskip=1pt,belowskip=3pt}
    \centering
    \subcaptionbox*{Reference ($a=0$)}{\includegraphics[width=0.325\linewidth]{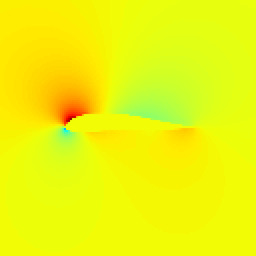}}\\
    \subcaptionbox*{$a=-0.16$}        {\includegraphics[width=0.325\linewidth]{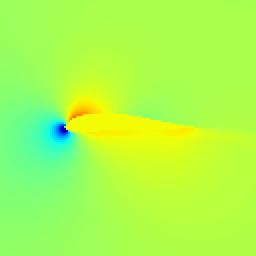}}
    \subcaptionbox*{$a=-0.08$}        {\includegraphics[width=0.325\linewidth]{figs/exp.airfoil/vely2_0012}}
    \subcaptionbox*{$a=-0.04$}        {\includegraphics[width=0.325\linewidth]{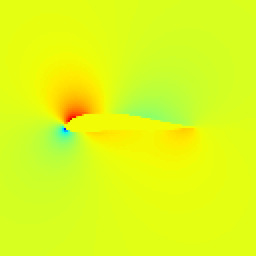}}\\
    \subcaptionbox*{$a=0.04$}         {\includegraphics[width=0.325\linewidth]{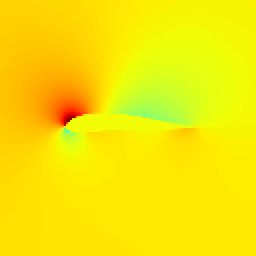}}
    \subcaptionbox*{$a=0.08$}         {\includegraphics[width=0.325\linewidth]{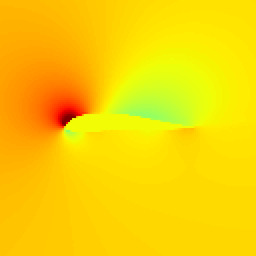}}
    \subcaptionbox*{$a=0.16$}         {\includegraphics[width=0.325\linewidth]{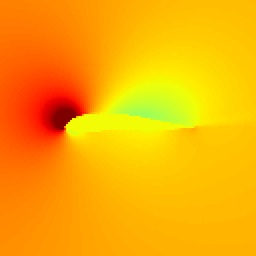}}
    \caption{Y-velocity}
    \label{fig:airfoil:vely}
  \end{subfigure}
  \caption{Airfoil Reynolds-averaged Navier-Stokes turbulence: An airfoil
    profile is simulated in two dimensions for different angles of attack (given
    as $a$ in radians below each image). These visualizations show the pressure
    distribution around the airfoil and the x and y components of the flow
    velocity (from top to bottom). In this case, the zero angle, $a=0$, is given
    as reference data set in the user studies.}
  \label{fig:airfoil:all}
\end{figure}

\clearpage

\begin{figure}
  \captionsetup[subfigure]{aboveskip=1pt,belowskip=6pt}
  \centering
  \begin{subfigure}[b]{\linewidth}
    \def\svgwidth{\linewidth}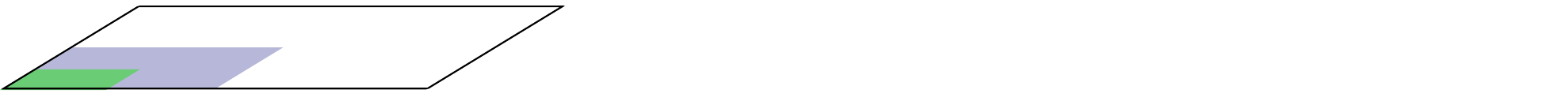
    \caption{Quadtree structure}
    \label{fig:quadtree}
  \end{subfigure}
  \begin{subfigure}[b]{\linewidth}
    \captionsetup[subfigure]{aboveskip=1pt,belowskip=3pt}
    \subcaptionbox*{$I^\mathrm{dv}_1$}{\includegraphics[width=0.245\linewidth]{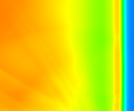}}
    \subcaptionbox*{$I^\mathrm{dv}_2$}{\includegraphics[width=0.245\linewidth]{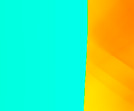}}
    \subcaptionbox*{$I^\mathrm{dv}_3$}{\includegraphics[width=0.245\linewidth]{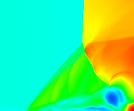}}
    \subcaptionbox*{$I^\mathrm{dv}_4$}{\includegraphics[width=0.245\linewidth]{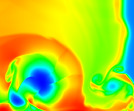}}
    \caption{Regions of the density value visualization in
      \myfigref{fig:vis-seven-methods:dv}}
    \label{fig:local:dv}
  \end{subfigure}
  \begin{subfigure}[b]{\linewidth}
    \captionsetup[subfigure]{aboveskip=1pt,belowskip=3pt}
    \subcaptionbox*{$I^\mathrm{gd}_1$}{\includegraphics[width=0.245\linewidth]{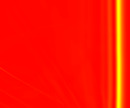}}
    \subcaptionbox*{$I^\mathrm{gd}_2$}{\includegraphics[width=0.245\linewidth]{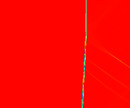}}
    \subcaptionbox*{$I^\mathrm{gd}_3$}{\includegraphics[width=0.245\linewidth]{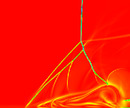}}
    \subcaptionbox*{$I^\mathrm{gd}_4$}{\includegraphics[width=0.245\linewidth]{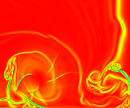}}
    \caption{Regions of the density gradient visualization in
      \myfigref{fig:vis-seven-methods:gd}}
    \label{fig:local:gd}
  \end{subfigure}
  \begin{subfigure}[b]{\linewidth}
    \captionsetup[subfigure]{aboveskip=1pt,belowskip=3pt}
    \subcaptionbox*{$I^\mathrm{dv}_{33}$}{\includegraphics[width=0.245\linewidth]{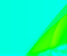}}
    \subcaptionbox*{$I^\mathrm{dv}_{34}$}{\includegraphics[width=0.245\linewidth]{figs/cut.vshock.dens/ref34}}
    \subcaptionbox*{$I^\mathrm{gd}_{33}$}{\includegraphics[width=0.245\linewidth]{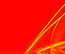}}
    \subcaptionbox*{$I^\mathrm{gd}_{34}$}{\includegraphics[width=0.245\linewidth]{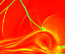}}
    \caption{Two bottom regions of $I^\mathrm{dv}_3$ (left) and
      $I^\mathrm{gd}_3$ (right)}
    \label{fig:local:2}
  \end{subfigure}
  \caption{Quadtree-based localization: From the root level $I$, the
    hierarchical subdivisions are denoted with subscripts according to the
    quadtree structure shown in (\subref{fig:quadtree}). The figures of
    (\subref{fig:local:dv}), (\subref{fig:local:gd}), and (\subref{fig:local:2})
    show examples of the localization for the reference data sets of the viscous
    shock tube simulations in \myfigref{fig:vis-seven-methods}. The superscripts
    ``dv'' and ``gd'' represent the density value visualization
    (\myfigref{fig:vis-seven-methods:dv}) and gradient visualization
    (\myfigref{fig:vis-seven-methods:gd}), respectively.}
  \label{fig:localization}
\end{figure}

\begin{figure}
  \captionsetup[subfigure]{aboveskip=1pt,belowskip=3pt}
  \centering
  \subcaptionbox*{$I^\mathrm{dv}_1$}   {\includegraphics[width=.24\linewidth]{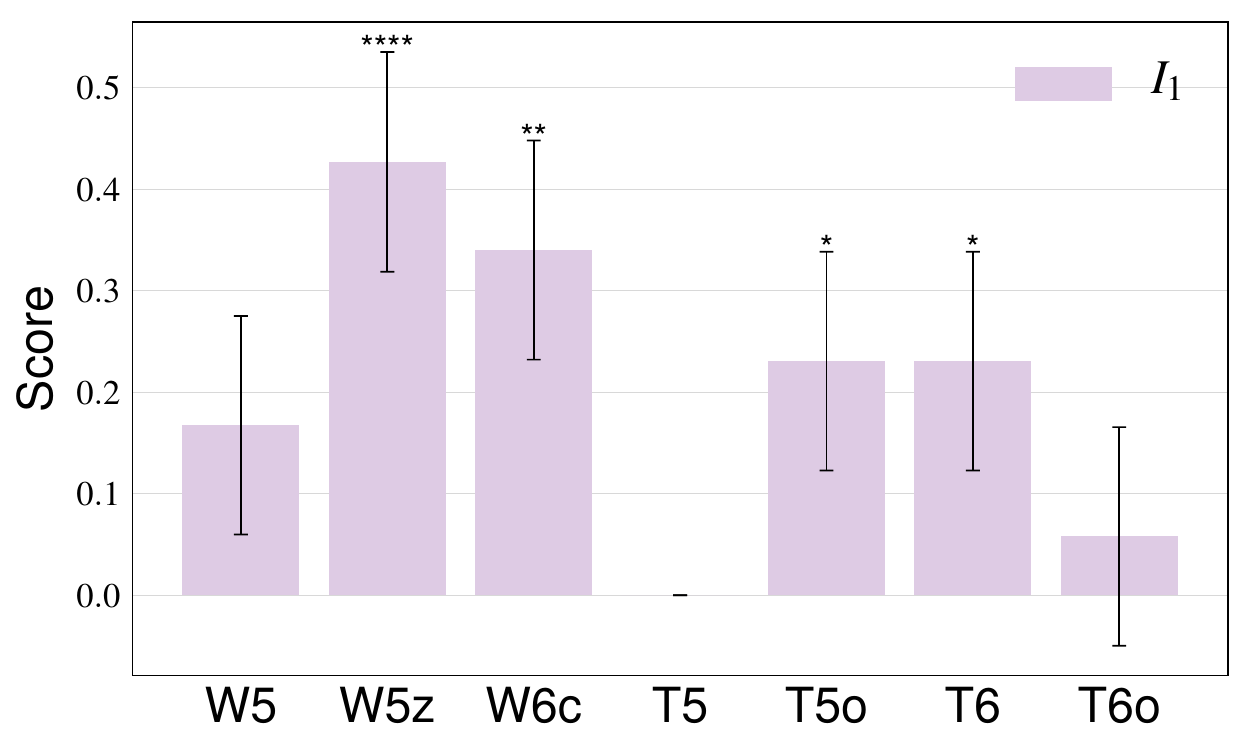}}
  \subcaptionbox*{$I^\mathrm{dv}_2$}   {\includegraphics[width=.24\linewidth]{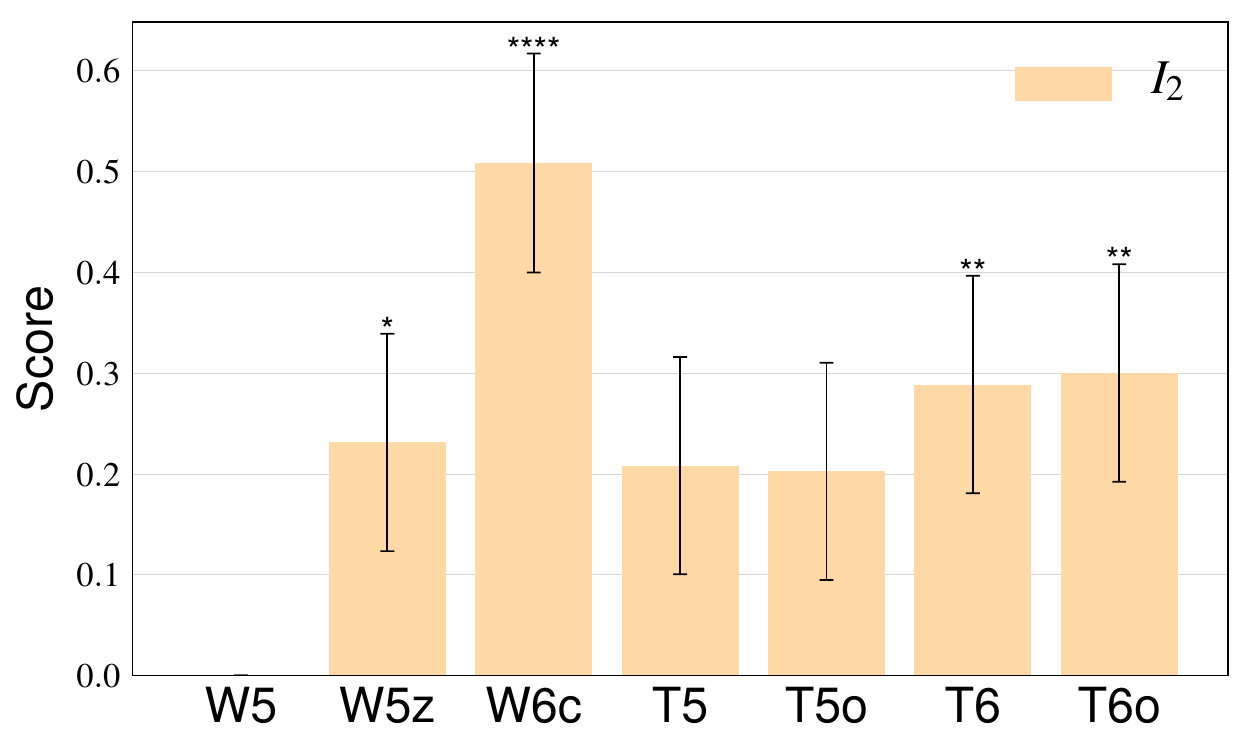}}
  \subcaptionbox*{$I^\mathrm{dv}_3$}   {\includegraphics[width=.24\linewidth]{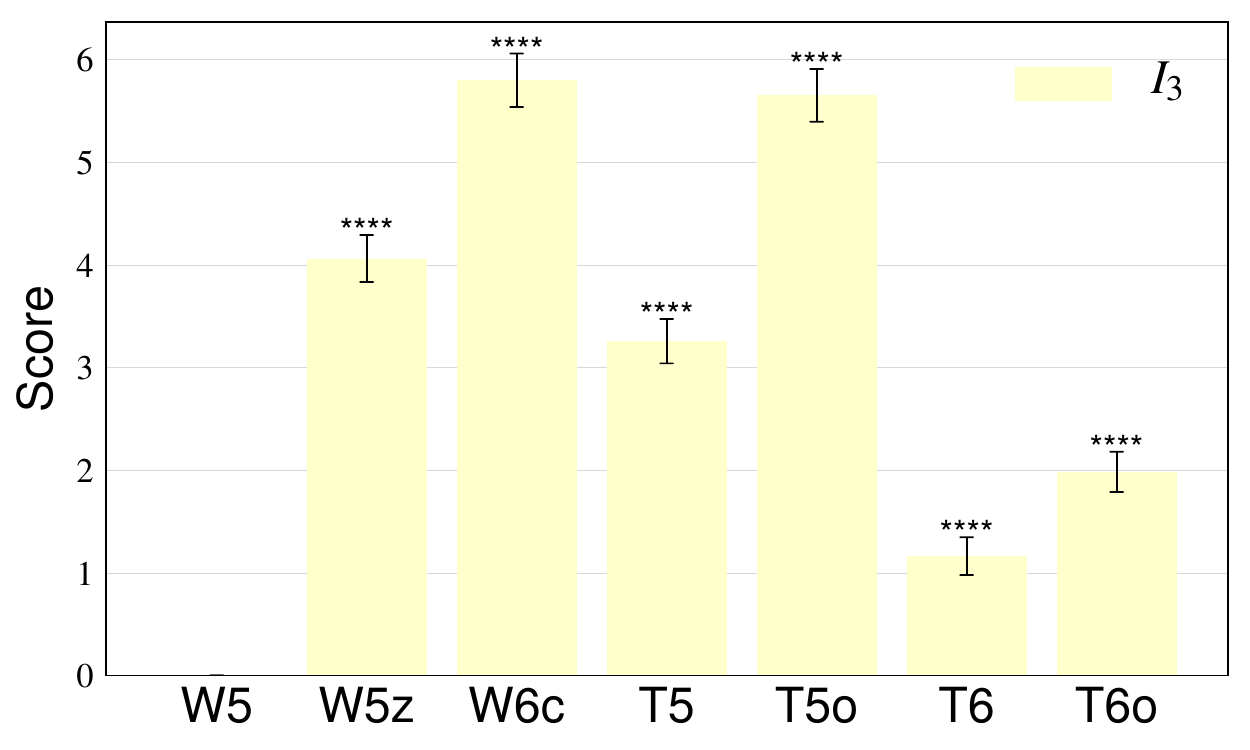}}
  \subcaptionbox*{$I^\mathrm{dv}_4$}   {\includegraphics[width=.24\linewidth]{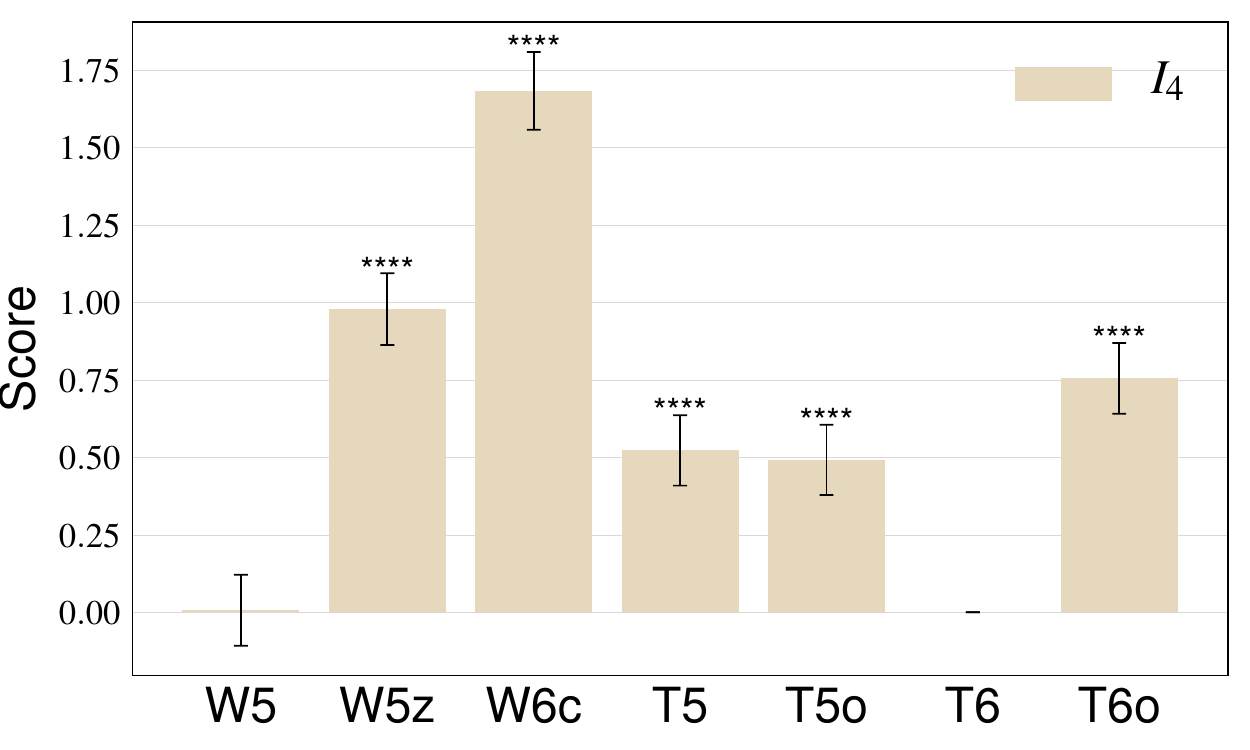}}\\
  \subcaptionbox*{$I^\mathrm{gd}_1$}   {\includegraphics[width=.24\linewidth]{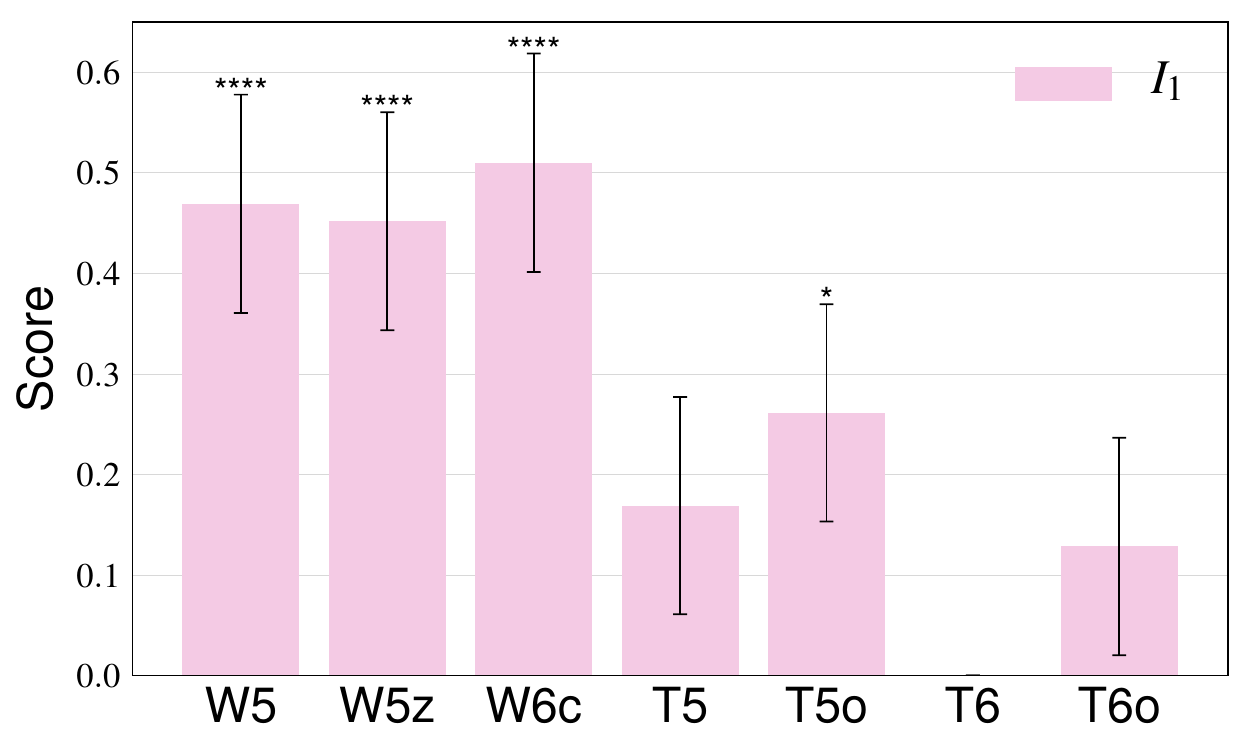}}
  \subcaptionbox*{$I^\mathrm{gd}_2$}   {\includegraphics[width=.24\linewidth]{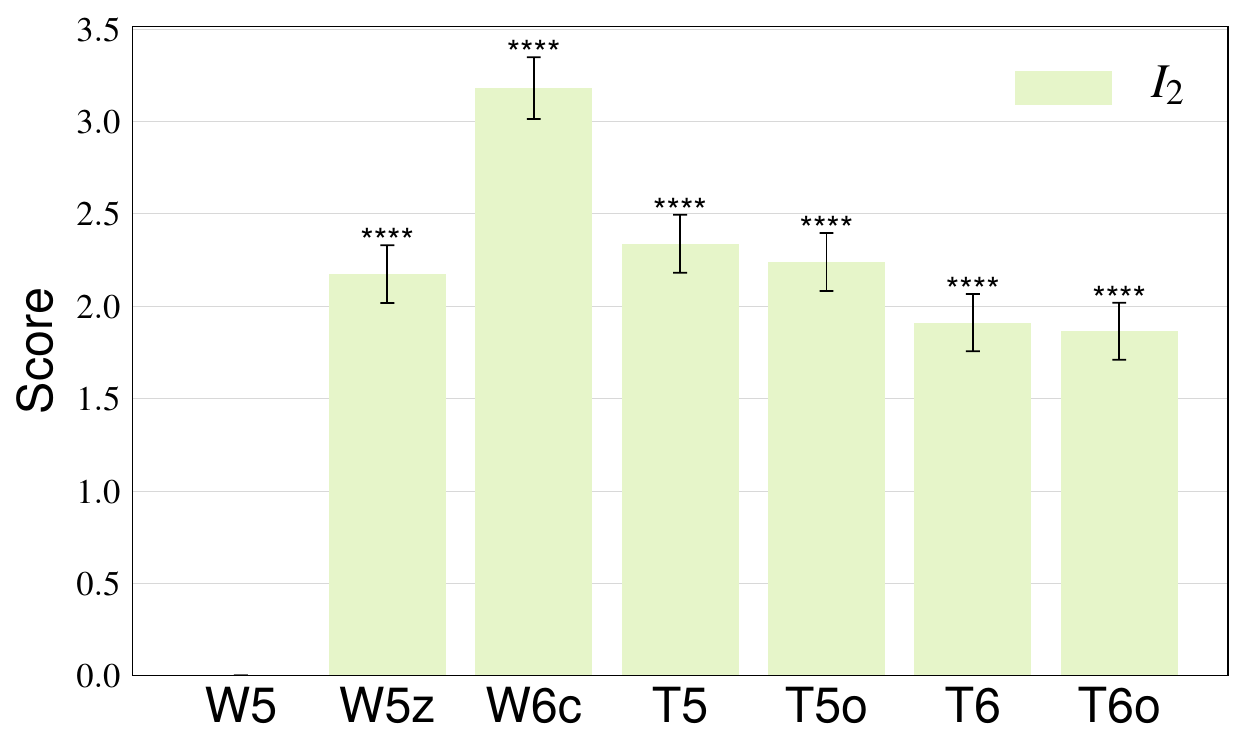}}
  \subcaptionbox*{$I^\mathrm{gd}_3$}   {\includegraphics[width=.24\linewidth]{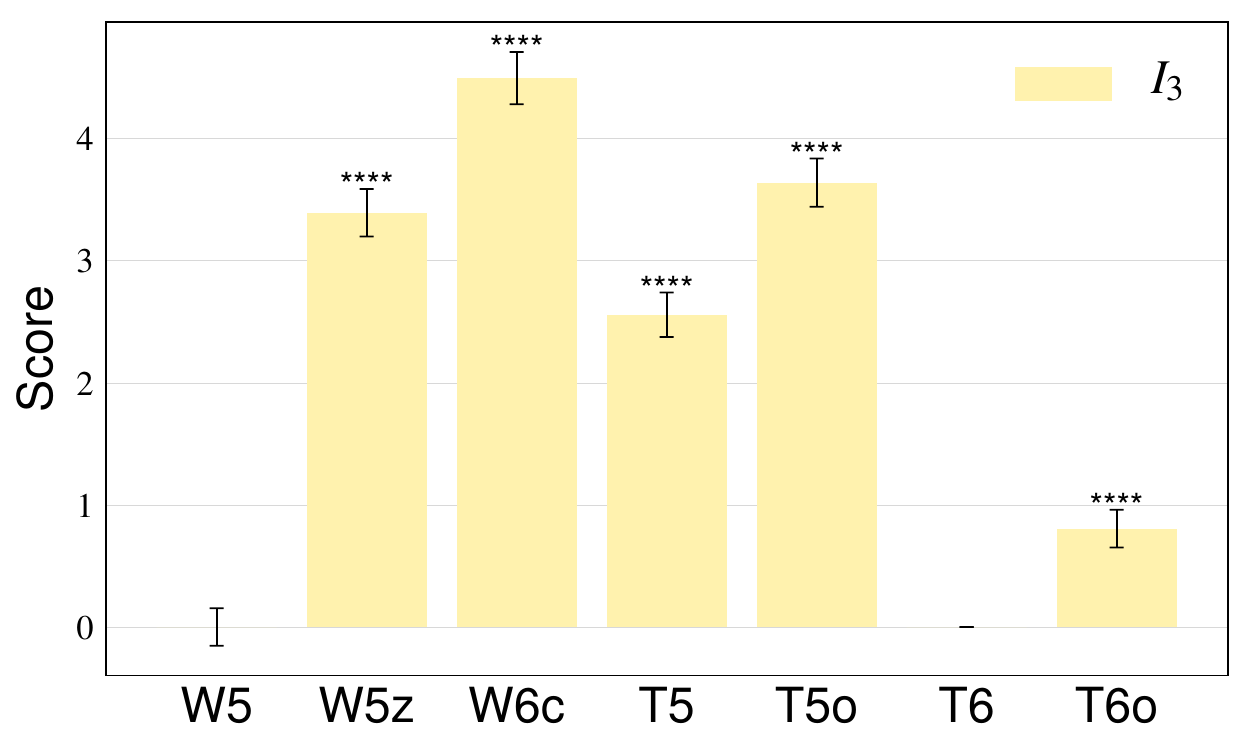}}
  \subcaptionbox*{$I^\mathrm{gd}_4$}   {\includegraphics[width=.24\linewidth]{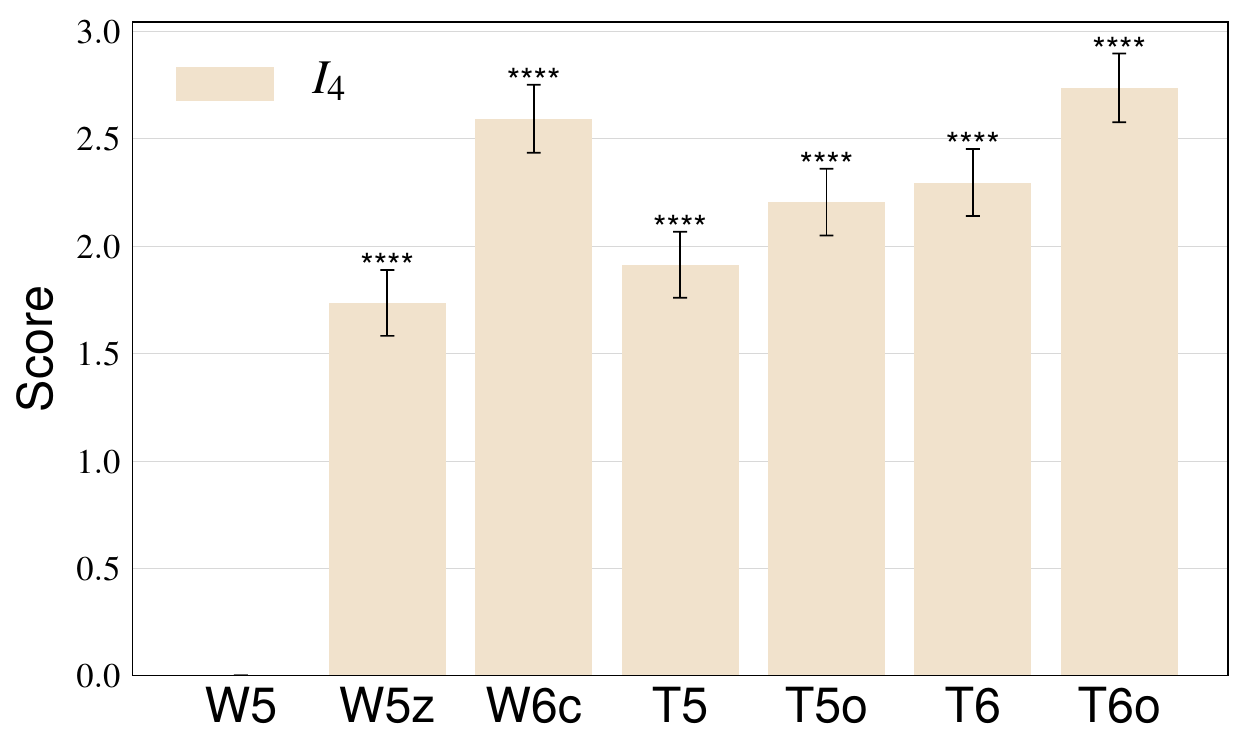}}\\
  \subcaptionbox*{$I^\mathrm{dv}_{33}$}{\includegraphics[width=.24\linewidth]{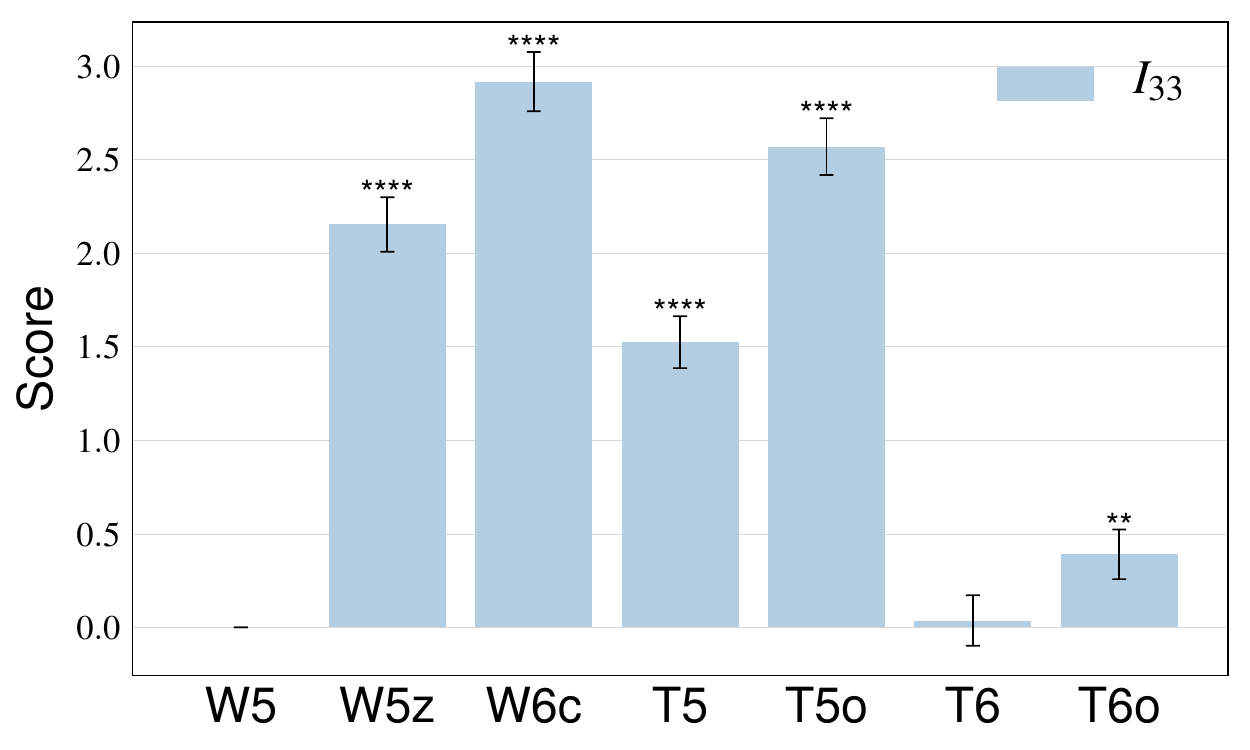}}
  \subcaptionbox*{$I^\mathrm{dv}_{34}$}{\includegraphics[width=.24\linewidth]{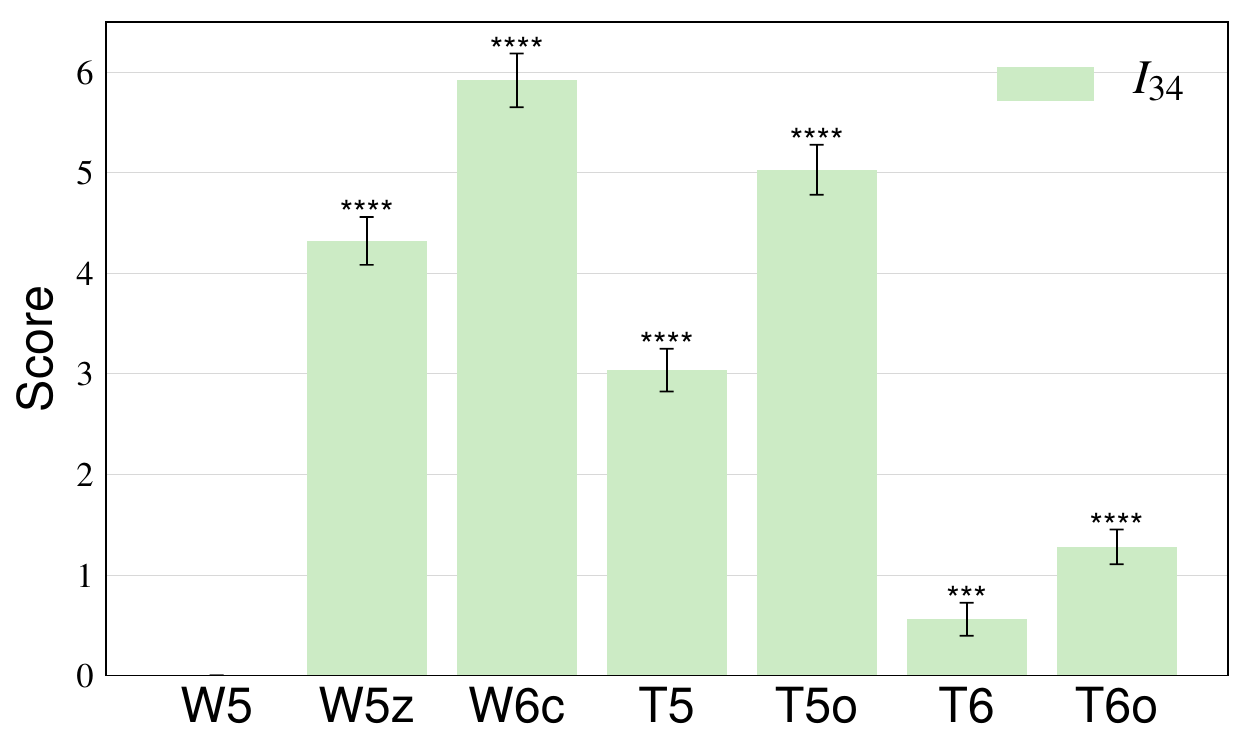}}
  \subcaptionbox*{$I^\mathrm{gd}_{33}$}{\includegraphics[width=.24\linewidth]{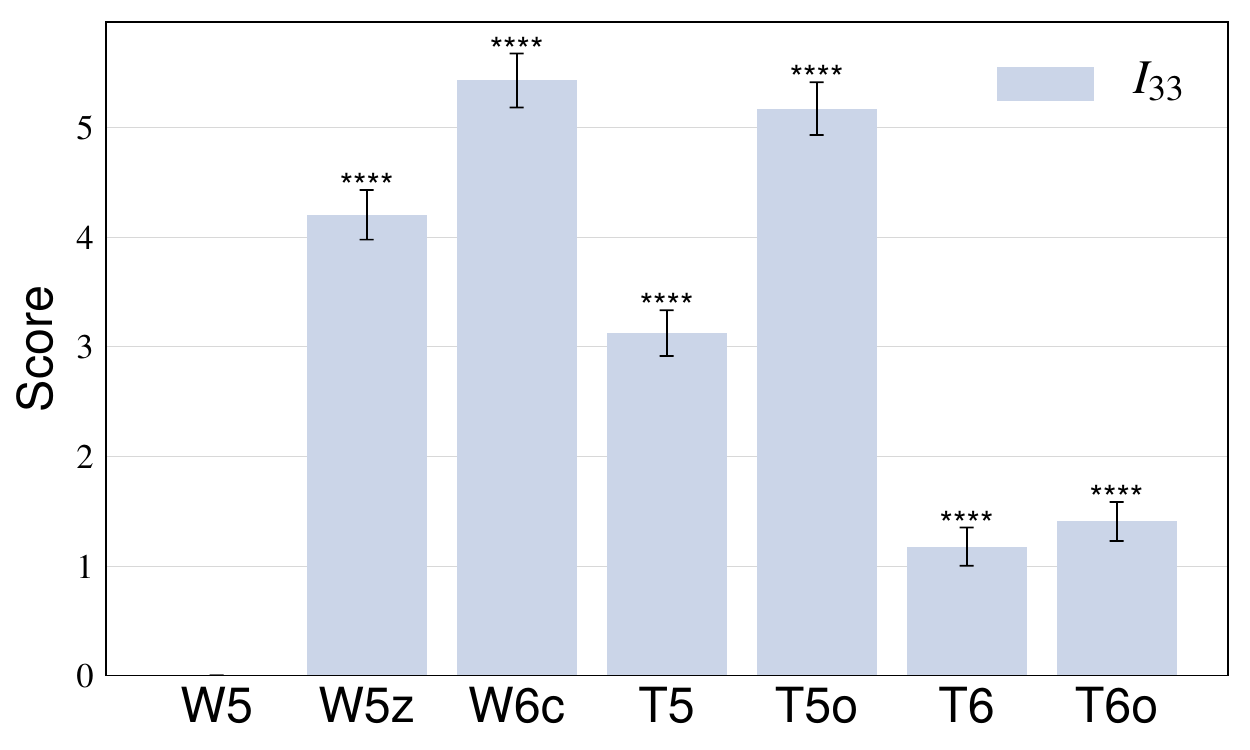}}
  \subcaptionbox*{$I^\mathrm{gd}_{34}$}{\includegraphics[width=.24\linewidth]{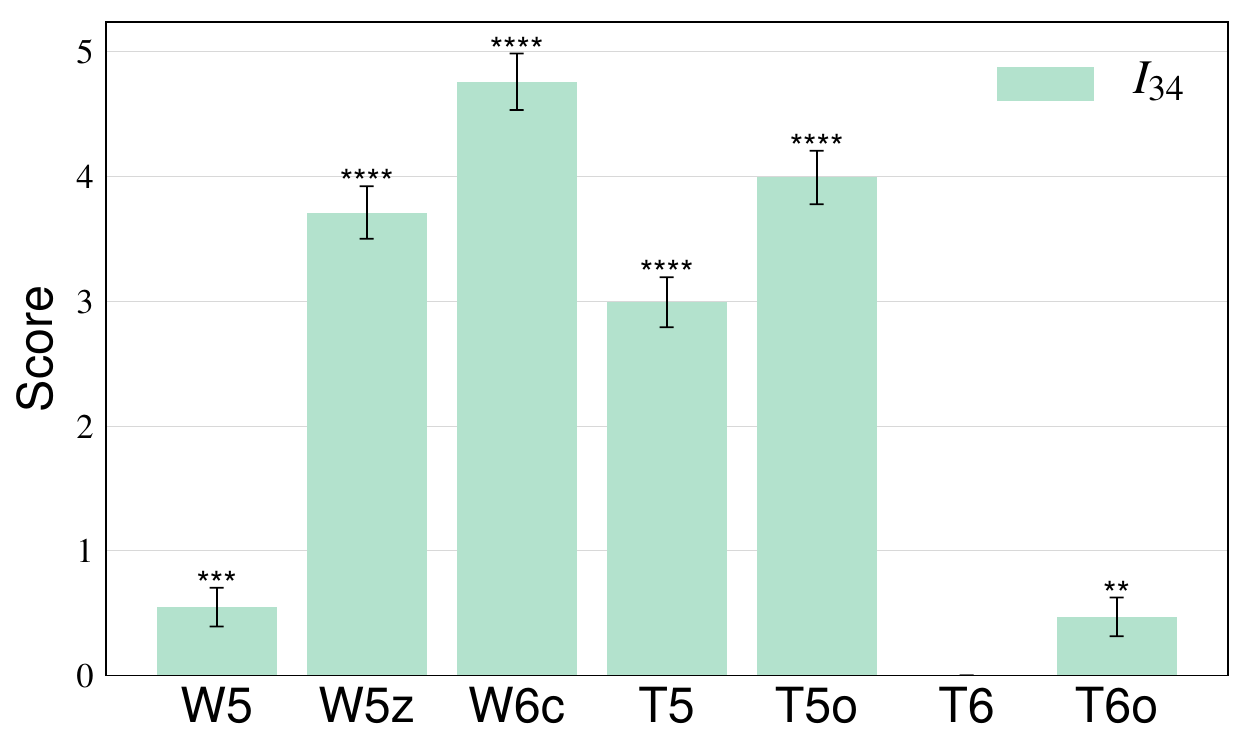}}
  \caption{User study results for the error localization (Experiment 3): The
    graphs show the performance scores of seven discretization schemes for the
    cutout regions of the viscous shock tube simulations. The superscripts
    ``dv'' and ``gd'' represent the density value visualizations
    (\myfigref{fig:vis-seven-methods:dv}) and gradient visualizations
    (\myfigref{fig:vis-seven-methods:gd}), respectively. The subscripts indicate
    each region according to \myfigref{fig:localization}. ****$P<0.0001$,
    **$P<0.01$, *$P<0.05.$}
  \label{fig:study:cuts}
\end{figure}

\clearpage

\begin{figure}
  \centering
  \begin{subfigure}[b]{\linewidth}
    \captionsetup[subfigure]{aboveskip=1pt,belowskip=3pt}
    \centering
    \subcaptionbox*{Reference}{\includegraphics[width=0.24\linewidth]{figs/exp.tgv/ref}}
    \subcaptionbox*{W5}    {\includegraphics[width=0.24\linewidth]{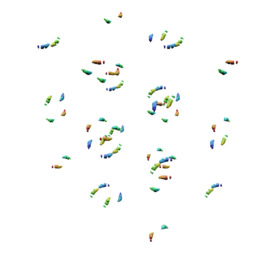}}
    \subcaptionbox*{W5z}   {\includegraphics[width=0.24\linewidth]{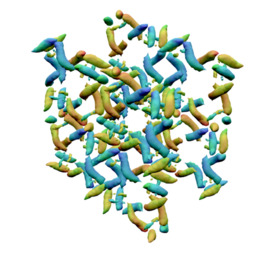}}
    \subcaptionbox*{W6c}   {\includegraphics[width=0.24\linewidth]{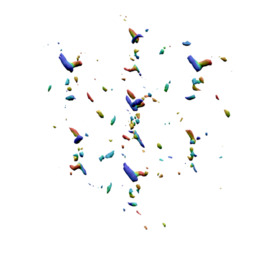}}\\
    \subcaptionbox*{T5}    {\includegraphics[width=0.24\linewidth]{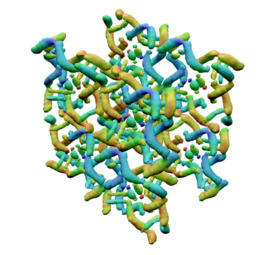}}
    \subcaptionbox*{T5o}   {\includegraphics[width=0.24\linewidth]{figs/exp.tgv/r064-teno5opt}}
    \subcaptionbox*{T6}    {\includegraphics[width=0.24\linewidth]{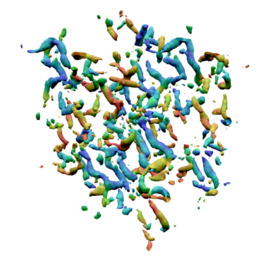}}
    \subcaptionbox*{T6o}   {\includegraphics[width=0.24\linewidth]{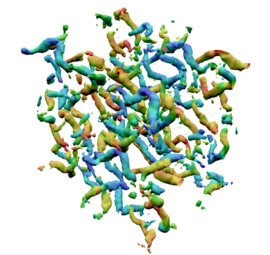}}\\
    \caption{$64^3$}
    \label{fig:tgv:064}
  \end{subfigure}
\end{figure}

\begin{figure}\ContinuedFloat
  \centering
  \begin{subfigure}[b]{\linewidth}
    \captionsetup[subfigure]{aboveskip=1pt,belowskip=3pt}
    \centering
    \subcaptionbox*{Reference}{\includegraphics[width=0.24\linewidth]{figs/exp.tgv/ref}}
    \subcaptionbox*{W5}    {\includegraphics[width=0.24\linewidth]{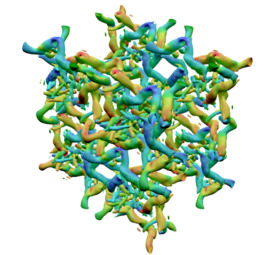}}
    \subcaptionbox*{W5z}   {\includegraphics[width=0.24\linewidth]{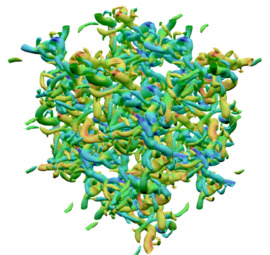}}
    \subcaptionbox*{W6c}   {\includegraphics[width=0.24\linewidth]{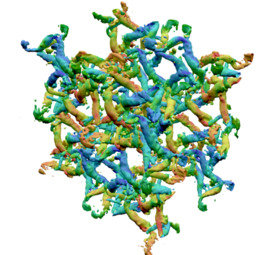}}\\
    \subcaptionbox*{T5}    {\includegraphics[width=0.24\linewidth]{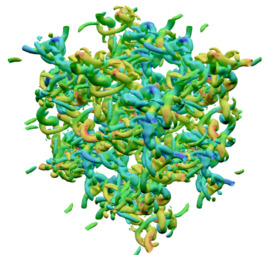}}
    \subcaptionbox*{T5o}   {\includegraphics[width=0.24\linewidth]{figs/exp.tgv/r128-teno5opt}}
    \subcaptionbox*{T6}    {\includegraphics[width=0.24\linewidth]{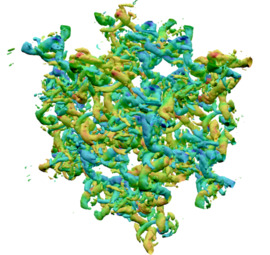}}
    \subcaptionbox*{T6o}   {\includegraphics[width=0.24\linewidth]{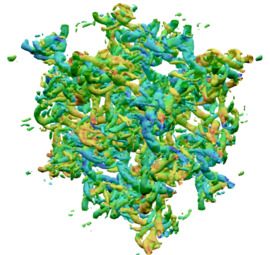}}\\
    \caption{$128^3$}
    \label{fig:tgv:128}
  \end{subfigure}  
\end{figure}

\begin{figure}\ContinuedFloat
  \centering
  \begin{subfigure}[b]{\linewidth}
    \captionsetup[subfigure]{aboveskip=1pt,belowskip=3pt}
    \centering
    \subcaptionbox*{Reference}{\includegraphics[width=0.24\linewidth]{figs/exp.tgv/ref}}
    \subcaptionbox*{W5}    {\includegraphics[width=0.24\linewidth]{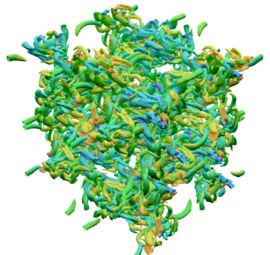}}
    \subcaptionbox*{W5z}   {\includegraphics[width=0.24\linewidth]{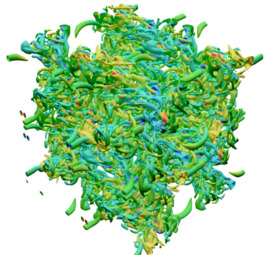}}
    \subcaptionbox*{W6c}   {\includegraphics[width=0.24\linewidth]{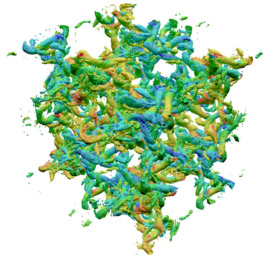}}\\
    \subcaptionbox*{T5}    {\includegraphics[width=0.24\linewidth]{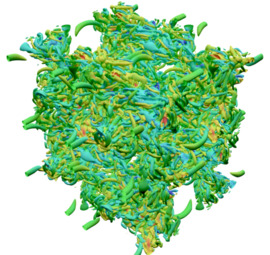}}
    \subcaptionbox*{T5o}   {\includegraphics[width=0.24\linewidth]{figs/exp.tgv/r256-teno5opt}}
    \subcaptionbox*{T6}    {\includegraphics[width=0.24\linewidth]{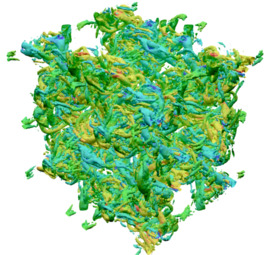}}
    \subcaptionbox*{T6o}   {\includegraphics[width=0.24\linewidth]{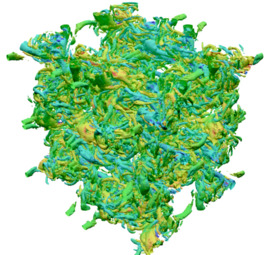}}\\
    \caption{$256^3$}
    \label{fig:tgv:256}
  \end{subfigure}
  \caption{\acl{tgv} flow simulations with different resolutions: The
    images visualize Q-criterion isosurfaces for $Q=3$ and are colored by
    $x$-component of vorticity.}
  \label{fig:tgv:all}
\end{figure}

\clearpage

\begin{figure}
  \centering
  \begin{subfigure}[b]{\linewidth}
    \captionsetup[subfigure]{aboveskip=1pt,belowskip=3pt}
    \centering
    \subcaptionbox*{Reference}{\includegraphics[width=0.24\linewidth]{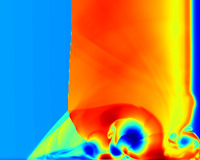}}
    \subcaptionbox*{W5}    {\includegraphics[width=0.24\linewidth]{figs/crossres.vshock.jet/x0320-weno}}
    \subcaptionbox*{W5z}   {\includegraphics[width=0.24\linewidth]{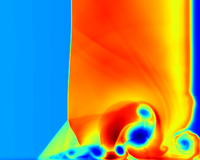}}
    \subcaptionbox*{W6c}   {\includegraphics[width=0.24\linewidth]{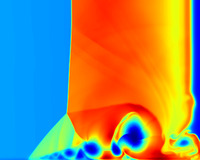}}\\
    \subcaptionbox*{T5}    {\includegraphics[width=0.24\linewidth]{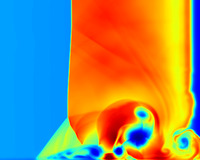}}
    \subcaptionbox*{T5o}   {\includegraphics[width=0.24\linewidth]{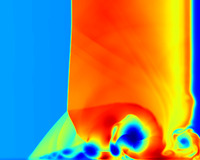}}
    \subcaptionbox*{T6}    {\includegraphics[width=0.24\linewidth]{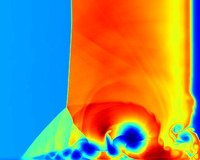}}
    \subcaptionbox*{T6o}   {\includegraphics[width=0.24\linewidth]{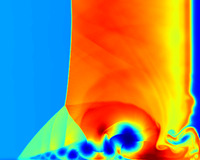}}\\
    \caption{$640\times320$}
    \label{fig:vshock:0320}
  \end{subfigure}
\end{figure}

\begin{figure}\ContinuedFloat
  \centering
  \begin{subfigure}[b]{\linewidth}
    \captionsetup[subfigure]{aboveskip=1pt,belowskip=3pt}
    \centering
    \subcaptionbox*{Reference}{\includegraphics[width=0.24\linewidth]{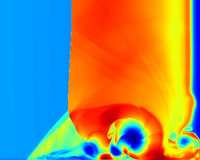}}
    \subcaptionbox*{W5}    {\includegraphics[width=0.24\linewidth]{figs/crossres.vshock.jet/x0480-weno}}
    \subcaptionbox*{W5z}   {\includegraphics[width=0.24\linewidth]{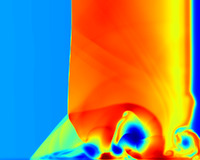}}
    \subcaptionbox*{W6c}   {\includegraphics[width=0.24\linewidth]{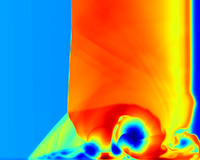}}\\
    \subcaptionbox*{T5}    {\includegraphics[width=0.24\linewidth]{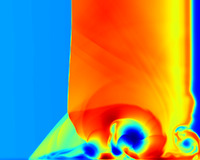}}
    \subcaptionbox*{T5o}   {\includegraphics[width=0.24\linewidth]{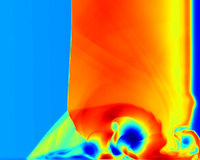}}
    \subcaptionbox*{T6}    {\includegraphics[width=0.24\linewidth]{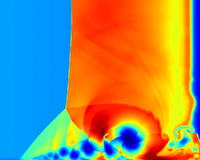}}
    \subcaptionbox*{T6o}   {\includegraphics[width=0.24\linewidth]{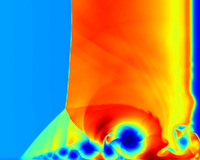}}\\
    \caption{$960\times480$}
    \label{fig:vshock:0480}
  \end{subfigure}
\end{figure}

\begin{figure}\ContinuedFloat
  \centering
  \begin{subfigure}[b]{\linewidth}
    \captionsetup[subfigure]{aboveskip=1pt,belowskip=3pt}
    \centering
    \subcaptionbox*{Reference}{\includegraphics[width=0.24\linewidth]{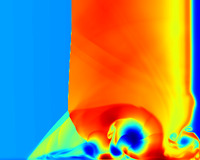}}
    \subcaptionbox*{W5}    {\includegraphics[width=0.24\linewidth]{figs/crossres.vshock.jet/x0640-weno}}
    \subcaptionbox*{W5z}   {\includegraphics[width=0.24\linewidth]{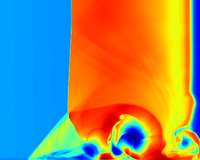}}
    \subcaptionbox*{W6c}   {\includegraphics[width=0.24\linewidth]{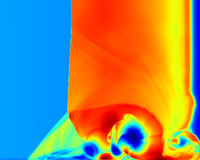}}\\
    \subcaptionbox*{T5}    {\includegraphics[width=0.24\linewidth]{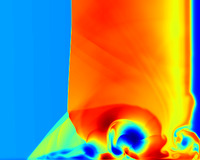}}
    \subcaptionbox*{T5o}   {\includegraphics[width=0.24\linewidth]{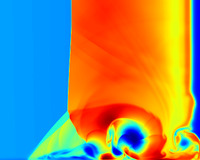}}
    \subcaptionbox*{T6}    {\includegraphics[width=0.24\linewidth]{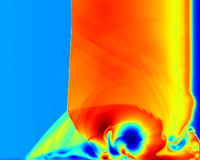}}
    \subcaptionbox*{T6o}   {\includegraphics[width=0.24\linewidth]{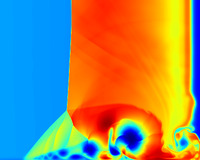}}\\
    \caption{$1280\times640$}
    \label{fig:vshock:0640}
  \end{subfigure}
\end{figure}

\begin{figure}\ContinuedFloat
  \centering
  \begin{subfigure}[b]{\linewidth}
    \captionsetup[subfigure]{aboveskip=1pt,belowskip=3pt}
    \centering
    \subcaptionbox*{Reference}{\includegraphics[width=0.24\linewidth]{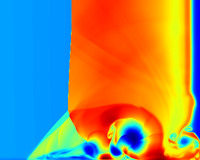}}
    \subcaptionbox*{W5}    {\includegraphics[width=0.24\linewidth]{figs/crossres.vshock.jet/x0960-weno}}
    \subcaptionbox*{W5z}   {\includegraphics[width=0.24\linewidth]{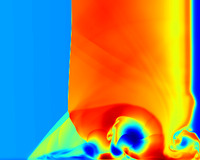}}
    \subcaptionbox*{W6c}   {\includegraphics[width=0.24\linewidth]{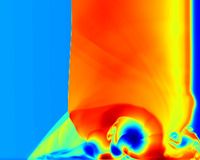}}\\
    \subcaptionbox*{T5}    {\includegraphics[width=0.24\linewidth]{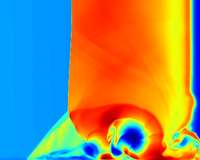}}
    \subcaptionbox*{T5o}   {\includegraphics[width=0.24\linewidth]{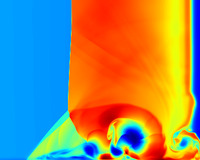}}
    \subcaptionbox*{T6}    {\includegraphics[width=0.24\linewidth]{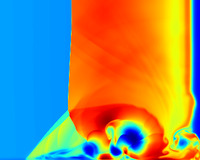}}
    \subcaptionbox*{T6o}   {\includegraphics[width=0.24\linewidth]{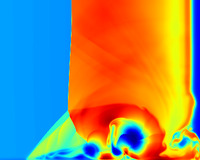}}\\
    \caption{$1920\times960$}
    \label{fig:vshock:0960}
  \end{subfigure}
\end{figure}

\begin{figure}\ContinuedFloat
  \centering
  \begin{subfigure}[b]{\linewidth}
    \captionsetup[subfigure]{aboveskip=1pt,belowskip=3pt}
    \centering
    \subcaptionbox*{Reference}{\includegraphics[width=0.24\linewidth]{figs/crossres.vshock.jet/x1280-ref}}
    \subcaptionbox*{W5}    {\includegraphics[width=0.24\linewidth]{figs/crossres.vshock.jet/x1280-weno}}
    \subcaptionbox*{W5z}   {\includegraphics[width=0.24\linewidth]{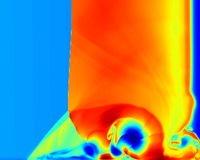}}
    \subcaptionbox*{W6c}   {\includegraphics[width=0.24\linewidth]{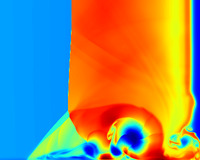}}\\
    \subcaptionbox*{T5}    {\includegraphics[width=0.24\linewidth]{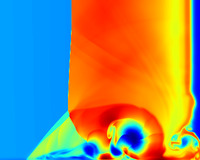}}
    \subcaptionbox*{T5o}   {\includegraphics[width=0.24\linewidth]{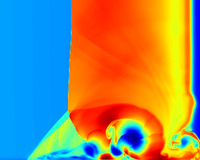}}
    \subcaptionbox*{T6}    {\includegraphics[width=0.24\linewidth]{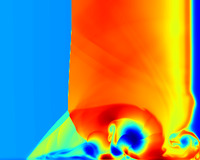}}
    \subcaptionbox*{T6o}   {\includegraphics[width=0.24\linewidth]{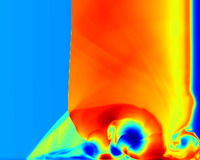}}\\
    \caption{$2560\times1280$}
    \label{fig:vshock:1280}
  \end{subfigure}
  \caption{Viscous shock tube simulations using different discretization schemes
    with different resolutions: The resolution of the reference data set is
    5120$\times$2560.}
    \label{fig:vshock:all}
\end{figure}

\FloatBarrier

\end{document}